\documentclass[journal,a4paper]{IEEEtran}
\usepackage{graphicx,amsmath,epsfig,amssymb,amsthm,latexsym,mathrsfs,epsf,enumerate,bm,color,epstopdf}
\newsavebox{\fminibox}
\newlength{\fminilength}

  \def\+{^\dagger}
\def\nequiv{\not\kern-.05em\equiv}
\def\egal{\kern-.5em=\kern-.5em}        % Moins d'espace autour de "="
\def\propt{\kern-.2em\propto\kern-.2em} % Idem
\def\argmin{\mathop{\mathrm{arg\,min}}} % car l'indice est reparti
\def\intdouble{\int\kern-0.3em\int}
\def\inttriple{\int\kern-0.3em\int\kern-0.3em\int}

\def\rond#1{\overset{\kern-0.33em~_\circ}{#1}}
\def\rondit[#1]#2{\overset{\kern#1~_\circ}{#2}}

\usepackage{subfloat}
\usepackage{float}
\usepackage{verbatim}
\usepackage{tikz}
\usetikzlibrary{shapes,arrows,shadows,chains}
\usepackage{tkz-graph}

\newtheorem{Theorem}{Theorem}

\newtheorem{CounterExample*}{$\overline{\hbox{\bf Example}}$}
\newtheorem{Definition}{Definition}

\newtheorem{Example*}{Example}

\newtheorem{Intuition*}{Intuition}
\newtheorem{Joke*}{Joke}
\newtheorem{Lemma}[Theorem]{Lemma}
\newtheorem{Lemma*}{Lemma}

\newtheorem{Note*}{Note}
\newtheorem{Open problem}{Open problem}

\newtheorem{Question*}{Question}
\newtheorem{Remark*}{Remark}

\newtheorem{Conjecture}{Conjecture}

\newcommand{\mcN}{{\mathcal N}}
\newcommand{\mcE}{{\mathcal E}}

\newcommand{\mcT}{{\mathcal T}}

\newcommand{\mcV}{{\mathcal V}}
\newcommand{\mcM}{{\mathcal M}}
\newcommand{\mcD}{{\mathcal D}}
\newcommand{\mcG}{{\mathcal G}}
\newcommand{\mcX}{{\mathcal X}}

\newcommand{\mbbR}{{\mathbb R}}

\newcommand{\uzm}{\underline{0}}
\newcommand{\uz}{\underline{{z}}}
\newcommand{\uw}{\underline{{w}}}
\newcommand{\ux}{\underline{{x}}}

\newcommand{\uW}{\underline{W}}
\newcommand{\uZ}{\underline{Z}}

\newcommand{\uX}{\underline{{X}}}

\newcommand{\bC}{{\mathbf{C}}}

\newcommand{\bSigma}{\mathbf{\Sigma}}

\newcommand{\bSigmaT}{\mathbf{\widetilde{\Sigma}}}

\newcommand{\bDelta}{\mathbf{\Delta}}

\newcommand{\bQ}{{\bf Q}}

\newcommand{\bI}{{\bf I}}
\newcommand{\bT}{{\bf T}}
\newcommand{\bL}{{\bf L}}
\newcommand{\bU}{{\bf U}}

\newcommand{\bP}{{\bf P}}

\begin{document}
%\title{A Greedy Approuch to Approximate Covariance Matrix of Renewable Generations with First Order Markov Chain by Minimizing Different Regret}
\sloppy

\title{Model Approximation Using Cascade of\\ Tree Decompositions}


%\author{\IEEEauthorblockN{-\\-\\-\\-\\-}
%\IEEEauthorblockA{-}}

\author{\IEEEauthorblockN{Navid Tafaghodi Khajavi and Anthony Kuh}
\IEEEauthorblockA{\\Deptartment of Electrical  Engineering,\\
University of Hawaii at Manoa, Honolulu, HI 96822\\
Email: navidt@hawaii$.$edu, kuh@hawaii$.$edu}}

%\and
%\IEEEauthorblockN{Toshihisa Tanaka}
%\IEEEauthorblockA{Dept. of Electrical and Electronic Engineering\\
%Tokyo University of Agriculture and Technology\\
%Tokyo, Japan\\
%Email: tanakat@cc$.$tuat$.$ac$.$jp}

\maketitle

\begin{abstract}
	
	We conduct a study of graphical models and discuss the quality of the statistical model approximation, especially when considering large graphs. 
	We are specifically looking at the statistical model approximation for jointly Gaussian random vectors. For Gaussian random vectors, one of the earliest model approximation methods is the covariance selection which is discussed by Dempster \cite{dempster}. Moreover, one of the simplest, widely used approximation models is a tree model that can be computed efficiently using the Chow-Liu algorithm \cite{chowliu}. This algorithm gives the optimal tree approximation for Gaussian distributions in the sense of the Kullback-Leibler (KL) divergence.
	
	In this paper, we present a general, multistage framework for graphical model approximation using a cascade of models such as trees. In particular, we look at the problem of covariance matrix approximation for Gaussian distributions as linear transformations of tree models. This is a new way to decompose the covariance matrix. Here, we propose an algorithm which incorporates the Cholesky factorization method to compute the decomposition matrix and thus can approximate a simple graphical model using a cascade of the Cholesky factorization of the tree approximation transformations. The Cholesky decomposition enables us to achieve a tree structure factor graph at each cascade stage of the algorithm which facilitates the use of the message passing algorithm since the approximated graph has less loops compared to the original graph. The overall graph is a cascade of factor graphs with each factor graph being a tree. This is a different perspective on the approximation model, and algorithms such as Gaussian belief propagation can be used on this overall graph. Here, we present theoretical result that guarantees the convergence of the proposed model approximation using the cascade of tree decompositions. In the simulations, we look at synthetic and real data and measure the performance of the proposed framework by comparing the KL divergences.
	
\end{abstract}

\section{Introduction}
\label{sec:intro}

Learning from high dimensional data requires large computational power which is not always available. The complexity limitation for different applications force us to compromise between the accuracy of the learning algorithm and its time complexity by using the best possible approximation algorithm subject to constraints on the graphical structure. There are many approximation algorithms that are proposed to impose structure given data. In signal processing and machine learning a fundamental problem is to balance performance quality (i.e. minimizing cost function) with computational complexity. A powerful tool in order to address this trade-off is graphical model selection. Model selection methods provide approximated models with desired accuracy as needed for different applications. Given data, different model selection algorithms impose different structure to model data \cite{dauwels_gm}.

There are methods in the literature such as the Chow-Liu minimum spanning tree (MST) \cite{chowliu}, the first order Markov chain approximation \cite{ISIT2013} and penalized likelihood methods such as LASSO \cite{LASSO} that can be used to approximate the correlation matrix with a more sparse graph while retaining good accuracy. 
The Chow-Liu MST algorithm for Gaussian distribution is to find the optimal tree structure using a Kullback-Leibler (KL) divergence cost function \cite{dempster}. The Chow-Liu algorithm utilizes the Kruskal algorithm \cite{kruskal}.
The first order Markov chain approximation uses a regret cost function to output a chain structured graph \cite{ISIT2013}. 
Penalized likelihood methods simplify the graph representation by eliminating some of the edges from the graphical structure, or in other words, imposing a sparsity condition on the inverse covariance matrix that represents the graphical model.
Recently, a tree approximation in a linear, underdetermined model is proposed in \cite{ICASSP17} where the solution is based on expectation, maximization (EM) algorithm combined with the Chow Liu algorithm.

Tree approximation algorithms are among the algorithms that reduce the number of computations in order to achieve quicker approximate solutions to a variety of problems. The tree approximations are made as it is much simpler to perform inference and estimation on trees rather than graphs that have cycles or loops. An example is applying the Gaussian belief propagation (BP) algorithm \cite{BP} which will converge to the maximum likelihood solution over loop-free graphs.
\footnote{The convergence of Gaussian BP with multiple loops is analyzed in literature \cite{BPcorrectness}.}
While these algorithms approximate the correlation matrix with a more sparse graph, in many cases as the number of nodes increases in large datasets, they fail to retain the desired accuracy \cite{MQ16ArXiv}. As a result, in many applications, we need to go beyond the tree structure approximation to achieve design accuracy that can be translated to any model approximation that can achieve a KL divergence bellow a certain design threshold.

Another related and mature body of work in literature is on mixture models \cite{jordanLGM} and \cite{kollerPGM}, including works on mixtures of tree approximations \cite{anandkumarMT12}, \cite{santanaMT} and \cite{meilaLMT} and Gaussian mixture model (GMM) \cite{GMM90} for graphical models.
While in this paper, we are generalizing a single-tree approximation algorithm and using a sequence of tree approximations for sparse model approximation, the aforementioned mixture of tree approximation methods consider parallel trees.

The purpose of this paper is to reduce the computation complexity of distributed algorithms in various applications while maintaining the desired approximation quality. To achieve this goal we approximate the associated Gaussian graphical model with a simpler, more tractable model. In this paper, we consider jointly Gaussian data and use cascade of tree transformation decompositions in order to perform model approximation for graphical models. The tree structure model is considered since this structure is simple and the optimal solution that minimizes the KL divergence can be easily computed using the Chow-Liu algorithm \cite{chowliu}. Furthermore, the tree structure model is a loop-free model and simplifies the implementation of distributed algorithms such as Gaussian BP. The cascade tree framework enables us to approximate a complex model with multiple stages of simple tractable models such as the tree structured model. We pick trees as the model and the Cholesky decomposition to factor the the tree structured covariance matrix at each stage of the cascade algorithm. Implementation of the Cholesky decomposition with the proper node ordering (permutation matrix) enables us to draw a {\it tree structured} factor graph for each step of the cascade tree decomposition transformation. This property facilitates the use of Gaussian BP algorithm over the aforementioned factor graph. We perform some simulations to confirm the results of this paper by looking at synthetic and real data and compare the performance of the proposed framework by comparing KL divergences. We also consider the singular value decomposition (SVD) and compare its performance to the Cholesky decomposition. Our simulation results also confirm the advantages of the cascade tree framework.

Many engineering and computer science applications require using graphs to model dependencies between nodes of the graph.  These applications include a diversity of areas from social networking to biomedical applications to transportation models to energy models.  For these applications graphs must be approximated by simpler structures to reduce computational complexity.  Here we briefly discuss an energy application.

{\it Smart grid applications:}
An electric distribution grid with measurements, distributed renewable energy sources, and decision making capabilities is referred to as the smart grid \cite{smartgrid}. Smart grids often have large number of states (e.g. node voltages) making it computationally inefficient to gather the data and perform central state estimation in real-time fashion. Moreover, central state estimation requires many communication links between sensors on the distribution grid resulting in large costs. In contrast, distributed state estimators can give reasonably good estimates for large power grid systems in real-time while decreasing the number of necessary communications links. This fact causes a trade off between calculation time and accuracy of estimation. The distributed state estimation method over a factor graph based on loopy Gaussian BP proposed in \cite{ISIT2013} can perform in real-time fashion. To assure the convergence of loopy Gaussian BP algorithm, \cite{ISIT2013} considers simple models of the distributed renewable energy sources by approximating covariance matrices of the distributed renewable energy sources with simpler covariance matrices that have tree-like structures. However tree-like structures  are poor approximates when the number of nodes is large \cite{MQ16ArXiv} leading us to consider more complex graphical structures.
Recently, \cite{dauwels_gm} also considered a BP approach for state estimation in power grid.

%{\it Big data applications:} I many real-world applications, extracting meaningful information from high dimensional data is a complex task. These applications can benefit from the cascade tree framework that we will present in this paper. One well-known example of these tasks is {\it Clustering} where we want to figure out the relationship between nodes/users and categorize them into clusters.

The rest of this paper is organized as follows. In section \ref{sec:tree_approx} we provide a summary of Gaussian tree approximation. The Gaussian model approximation as a transformation is also discussed in this section. Section \ref{sec:cascade} presents the theory behind the proposed model approximation framework. The  symmetric correlation approximation matrix (CAM) is defined and the convergence theorem is discussed in this section. Section \ref{sec:alg} provides a greedy algorithm for the model approximation using cascade of tree decompositions. The proposed algorithm is based on the symmetric CAM, the tree approximation algorithm and its Cholesky decomposition. This algorithm is suitable for message passing and Gaussian BP over factor graphs since we use the Cholesky decomposition at each of the cascade stages. In section \ref{sec:alg} we also present a simple example illustrating the Cholesky algorithm transformations and the cascade of factor graphs. Section \ref{sec:sim} provides some simulations over synthetic examples as well as a real solar data example from the island of Oahu obtained from NREL website and investigates the quality of the proposed model approximation by looking at the KL divergence. Finally, Section \ref{sec:con} summarizes results of this paper.

\paragraph*{Notation}
Upper case and lower case letters denote random variables and their realizations, respectively;
underlined letters stand for vectors; boldface upper case letters denote matrices;
$\left( \cdot \right)^{T}$, $\textnormal{tr}\{\cdot\}$ and $| \cdot |$ stand for transpose, trace and determinant of a matrix. In the rest of this paper, with shorthand notion, when we use the KL divergence between covariance matrices it means the KL divergence between their associated multivariate, zero-mean Gaussian distributions.

\section{Gaussian tree approximation}
\label{sec:tree_approx}

In this section, we first review the tree approximation algorithm for Gaussian distributions. Then we explain the framework for the covariance transformation decomposition for any given model such as the tree model. The tree structure is a simple graphical model and can be computed efficiently. The loop-free structure of the tree structure also facilitates the implementation of distributed algorithms such as Gaussian BP. Later in the next section, we use the cascade of tree transformation decompositions to perform model approximation for graphical models.

\subsection{Tree approximation for Gaussian distributions}
In the tree approximation, we want to approximate a multivariate distribution by the product of lower order component distributions \cite{lewis_approx}.
%For the purpose of tree approximation, the maximum order of these lower order distributions is two, i.e. no more than pairs of variables.
Let $\uX \sim \mcN (\uzm , \bSigma)$ (i.e. jointly Gaussian with mean 0 and covariance matrix $\bSigma$) where $\uX \in \mbbR^n$ have the graph representation $\mcG=(\mcV, \mcE)$ where sets $\mcV$ and $\mcE$ are the set of all vertices and edges of the graph representing $\uX$.\footnote{Here, we assume that all nodes are connected in the graphical structure of vector $\uX$.}
Let $\uX_{\mcT} \sim \mcN (\uzm , \bSigmaT)$ have the graph representation $\mcG_{\mcT}=(\mcV, \mcE_{\mcT})$ where $\mcE_{\mcT} \subseteq \mcE$ is a set of edges that represents a tree structure.
The joint probability density function can be represented by joint pdfs of two variables and marginal PDFs in the following convenient form
\begin{equation}
\label{eq:prod_approx}
f_{\uX_\mcT}(\ux) = \prod_{(u,v) \in \mcE_\mcT} \frac{f_{\uX^u,\uX^v}(\ux^u , \ux^v) }{f_{\uX^u}(\ux^u) f_{\uX^v}(\ux^v)} \prod_{o \in \mcV} f_{\uX^o}(\ux^o).
\end{equation}
\begin{Definition}
	Let $\mcT_{\bSigma}$ denote the set of all positive definite covariance matrices with following properties:
	\\
	1) These covariance matrices have tree structured Gaussian graphical models;
	\\
	2) Picking any covariance matrix in this set, $\bSigmaT \in \mcT_{\bSigma}$, the Gaussian distributions $\mcN (\uzm, \bSigmaT)$ and $\mcN (\uzm, \bSigma)$ have the same marginal distributions and joint distribution of two variables over the tree structured graph, $\mcG_\mcT$. \hfill \IEEEQED
\end{Definition}
%In other words, for all $\bSigmaT \in \mcT_{\bSigma}$, the corresponding zero-mean Gaussian distribution have the same marginals over the graph $\mcG_{\mcT}$ as the zero-mean Gaussian distribution with covariance matrix $\bSigma$. 
In the above definition $\mcN (\uzm, \bSigmaT)$ obeys the product rule given in \eqref{eq:prod_approx}. Also, note that, the cardinality of the set $\mcT_{\bSigma}$ is finite \cite{cayley} since the number of all possible tree structured graphs with $n$ nodes is finite.

\begin{Definition}
	The KL divergence between two multivariate continuous distributions $p_{\uX}(\uX)$ and $q_{\uX}(\uX)$ is defined as
	$$\mcD \left( p_{\uX}(\ux)||q_{\uX}(\ux) \right) = \int_{\mcX} p_{\uX}(\ux) \log \frac{p_{\uX}(\ux)}{q_{\uX}(\ux)} \; d \ux$$
	where $\mcX$ is the feasible set. \hfill \ensuremath{\blacksquare}
\end{Definition}

Chow-Liu MST method \cite{chowliu}, was initially proposed for approximating the joint distribution of discrete variables by product of lower order distributions similar to \eqref{eq:prod_approx} which involves no more than the pair of variables. The proposed KL divergence is used to quantify the distance between any distribution and its tree structure approximation. 

The Chow-Liu MST algorithm for Gaussian distributions, minimizes the following optimization problem in order to find the optimal tree structured covariance matrix, $\bSigma_\mcT \in \mcT_\bSigma$
\begin{equation}
\label{eq:chowliuopt_givenDist}
\bSigma_\mcT = \argmin_{\bSigmaT \in \mcT_\bSigma} \; \mcD ( f_{\uX}(\ux) || f_{\uX_\mcT}(\ux) ).
\end{equation}
Here, $\mathcal{D}^\star \triangleq - \frac{1}{2} \textnormal{log} (|\bSigma\bSigma_\mcT^{-1}|)$ is minimum the KL divergence that gives the distance between the given distribution and its optimal tree approximation.
It is shown in \cite{chowliu} that the optimal solution for this problem \ref{eq:chowliuopt_givenDist} can be found efficiently using greedy algorithms \cite{kruskal}, \cite{prim}. Their algorithm can be easily generalized for approximating the optimal tree structure of the joint distribution of Gaussian variables using equation \eqref{eq:prod_approx} by adding edges one at a time \cite{Lbanded_kavcic}.
In other words, given the knowledge of $\bSigma$, the Chow-Liu algorithm can efficiently compute the optimal solution, i.e. $\bSigma_\mcT = \text{chowliu}(\bSigma)$.

\noindent {\bf Remark:} In case that the covariance matrix $\bSigma$ is not available, we can replace it with the empirical covariance matrix obtained from data.
%
%\textcolor{red}{\subsection{Reverse KL divergence for Gaussian tree approximation}}
%\vspace{5cm}

\subsection{Gaussian model approximation as a transformation}
Any zero-mean multivariate Gaussian distribution such as $\mcN(\uzm, \bSigma)$ where $\bSigma$ is the covariance matrix, can be obtained through a linear transformation of the multivariate standard normal distribution, $\mcN(\uzm, \bI)$ (figure \ref{fig:t1}) where $\bI$ is the identity matrix. Moreover, the decomposition matrix $\bC$ is defined as a square matrix that factors the covariance matrix $\bSigma$, i.e. $\bSigma \triangleq \bC \bC^T$. In this scenario, the decomposition matrix $\bC$ is also the transformation matrix. We focus on the decomposition matrix $\bC$ in more detail in section \ref{sec:alg}, some of the possible matrix decompositions that can be used to efficiently compute $\bC$ are the Cholesky decomposition and singular value decomposition (SVD). Let's assume that the desired model covariance matrix, $\bSigma_\mcM$ and its decomposition matrix, $\bC_\mcM$, i.e. $\bSigma_\mcM \triangleq \bC_\mcM \bC_\mcM^T$, are given. Then, from figure \ref{fig:t2}, the model distribution $\mcN(\uzm, \bSigma_\mcM)$ is the transformation of the multivariate standard normal distribution. However, to generate the Gaussian distribution with covariance matrix, $\mcN(\uzm, \bSigma)$, using the model decomposition matrix, $\bC_\mcM$, the input distribution, $\mcN(\uzm, \bSigma)$, has to have a certain covariance matrix, $\bDelta$. This covariance matrix is called the symmetric correlation approximation matrix and is defined as $\bDelta = \bC_\mcM^{-1} \bSigma \bC_\mcM^{-T}$. We will give a formal definition for the symmetric CAM in section \ref{sec:cascade} where we consider cascade of tree approximation decompositions for graphical model approximation.

\tikzstyle{dist}=[ text width=8 em, minimum height=10 mm,text centered,node distance=3 em]
\tikzstyle{block}=[draw,rectangle,  text width=2.5 em, minimum height=10 mm,text centered,node distance=3 em]
\tikzstyle{line}=[draw, -stealth,thick]
\begin{subfigures}
\begin{figure}
	\centering	
	\begin{tikzpicture}
	\node [dist] (s) {$\mcN(\uzm, \bI)$};
	\node [block, right of = s, xshift=5em] (b1) {$\bC$};
	\node [dist, right of = b1, xshift=5em] (e) {$\mcN(\uzm, \bSigma)$};
	\path [line] (s) -- (b1);
	\path [line] (b1) -- (e);
	\end{tikzpicture}
	\caption{Transformation from $\mcN(\uzm, \bI)$ to $\mcN(\uzm, \bSigma)$ using decomposition of the covariance matrix, $\bSigma$.}
	\label{fig:t1}
%\end{figure}
\vspace{0.5cm}
%\begin{figure}
	\centering
	\begin{tikzpicture}
	\node [dist] (s) {$\mcN(\uzm, \bI)$};
	\node [block, right of = s, xshift=5em] (b1) {$\bC_\mcM$};
	\node [dist, right of = b1, xshift=5em] (e) {$\mcN(\uzm, \bSigma_{\mcM})$};
	\path [line] (s) -- (b1);
	\path [line] (b1) -- (e);
	\end{tikzpicture}
	\caption{Transformation from $\mcN(\uzm, \bI)$ to $\mcN(\uzm, \bSigma_\mcM)$ using decomposition of the model covariance matrix, $\bSigma_\mcM$.}
	\label{fig:t2}
%\end{figure}
\vspace{0.5cm}
%\begin{figure}
	\centering	
	\begin{tikzpicture}
	\node [dist] (s) {$\mcN(\uzm, \bDelta)$};
	\node [block, right of = s, xshift=5em] (b1) {$\bC_\mcM$};
	\node [dist, right of = b1, xshift=5em] (e) {$\mcN(\uzm, \bSigma)$};
	\path [line] (s) -- (b1);
	\path [line] (b1) -- (e);
	\end{tikzpicture}
	\caption{Transformation from $\mcN(\uzm, \bDelta)$ to $\mcN(\uzm, \bSigma)$ using decomposition of the model covariance matrix, $\bSigma_\mcM$.}
	\label{fig:t3}
\end{figure}
\end{subfigures}

\noindent{\bf Remark:} {\it Invariance of Gaussian KL divergence with respect to transformation.} The KL divergence between the input Gaussian distributions in figures \ref{fig:t2} and \ref{fig:t3} is invariant to the transformation $\bC_\mcM$, i.e. it is equal to the KL divergence between the output Gaussian distributions in figures \ref{fig:t2} and \ref{fig:t3}, ($\mcD( \bDelta || \bI) = \mcD( \bSigma || \bSigma_\mcM )$).

\noindent{\bf Remark:} In the rest of this paper we consider the tree approximation as our model.

\section{Model Approximation Using Cascade of Tree Decompositions Principle}
\label{sec:cascade}

\begin{subfigures}
	\begin{figure*}
		\centering	
		\begin{tikzpicture}
		\node [dist] (s) {$\uZ_i \sim \mcN(\uzm, \bDelta_i)$};
		\node [block, right of = s, xshift=5em] (b1) {$\bC_{\mcT_i}$};
		\node [dist, right of = b1, xshift=5em] (b2) {$\uZ_{i-1}\quad\;\;$ ... $\quad\;\;\uZ_1$};
		\node [block, right of = b2, xshift=5em] (b3) {$\bC_{\mcT_1}$};
		\node [dist, right of = b3, xshift=5em] (e) {$\uX \sim \mcN(\uzm, \bSigma)$};
		\path [line] (s) -- (b1);
		\path [line] (b1) -- (b2);
		\path [line] (b2) -- (b3);
		\path [line] (b3) -- (e);
		\end{tikzpicture}
		\caption{The $i$ stages of the model transformation from $\uZ_i \sim \mcN(\uzm, \bDelta_i)$ to $\uX \sim \mcN(\uzm, \bSigma)$ using cascade tree decompositions.}
		\label{fig:t4}
		%\end{figure*}
		\vspace{0.5cm}
		%\begin{figure*}
		\centering	
		\begin{tikzpicture}
		\node [dist] (s) {$\uW \sim \mcN(\uzm, \bI)$};
		\node [block, right of = s, xshift=5em] (b1) {$\bC_{\mcT_l}$};
		\node [dist, right of = b1, xshift=5em] (b2) {...};
		\node [block, right of = b2, xshift=5em] (b3) {$\bC_{\mcT_1}$};
		\node [dist, right of = b3, xshift=5em] (e) {$\uX_{\mcM_l}\!\!\sim \mcN(\uzm, \bSigma_{\mcM_l}\!)$};
		\path [line] (s) -- (b1);
		\path [line] (b1) -- (b2);
		\path [line] (b2) -- (b3);
		\path [line] (b3) -- (e);
		\end{tikzpicture}
		\caption{The $l$ stages of model approximation using cascade tree transformation decomposition framework. The model approximation is generated by passing $\uW \sim \mcN(\uzm, \bI)$ through the $l$ steps of cascade trees and is $\uX_{\mcM_l} \sim \mcN(\uzm, \bSigma_{\mcM_l})$.}
		\label{fig:t5}
	\end{figure*}
\end{subfigures}

In this section, we focus on the cascade of trees framework for model selection using the tree decomposition transformations. We formulate the problem by considering the tree approximation as a transformation and we use multiple stages of these cascade trees to do model approximation.
Let $\bSigma \triangleq \bC \bC^T$ and $\bSigma_\mcT \triangleq \bC_\mcT \bC_\mcT^T$ where $\bC$ and $\bC_\mcT$ are square transformation matrices that decompose the covariance matrices, $\bSigma$ and $\bSigma_\mcT$. Without loss of generality, in the rest of this paper, we look at the zero-mean Gaussian distributions with normalized covariance matrix $\bSigma$, i.e. covariance and correlation matrices are the same. Factoring covariances enable us to look at the problem as a transformation, as it is shown in figure 1. There are different decomposition algorithms to factor covariances such as the Cholesky decomposition and SVD. While we discuss the performance of the cascade of trees framework for model selection here, picking the decomposition algorithm will be discussed in section \ref{sec:alg}.

\begin{Definition}
	The symmetric correlation approximation matrix (CAM) for the tree approximation model is defined as $\bDelta \triangleq \bC_{\mcT}^{-1} \, \bSigma \, \bC_{\mcT}^{-T}$.  \hfill \IEEEQED
\end{Definition}

The symmetric CAM for each step of the cascade tree algorithm is also defined using the transformation matrix $\bC_{\mcT_i}$ and the previous step symmetric CAM.

\begin{Definition}
	The symmetric correlation approximation matrix for the $i$-th step of the cascade tree approximation is defined as $\bDelta_i \triangleq \bC_{\mcT_i}^{-1} \bDelta_{i-1} \bC_{\mcT_i}^{-T}$ where $\bDelta_0 \triangleq \bSigma$, $\bSigma_{\mcT_i} = \text{chowliu}(\bDelta_{i-1})$ and $\bSigma_{\mcT_i} \triangleq \bC_{\mcT_i} \bC_{\mcT_i}^T $ where $\bC_{\mcT_i}$ is the decomposition for the $i$-the step covariance matrix, $\bSigma_{\mcT_i}$.  \hfill \IEEEQED
\end{Definition}

Figures \ref{fig:t4} and \ref{fig:t5} show schematic diagrams associated with the cascade tree framework.
In figure \ref{fig:t4}, we want to model the zero-mean multivariate Gaussian distribution, $\uX \sim \mcN (\uzm , \bSigma)$, using the cascade of tree decomposition transformations. 
Let $\uX_{\mcT_i} \sim \mcN (\uzm , \bSigma_{\mcT_i})$ be the tree approximation distribution for the residue random vector $\uZ_{i-1} \sim \mcN (\uzm , \bDelta_{i-1})$\footnote{$\uZ_0 \sim \mcN (\uzm , \bSigma)$.} where $\bSigma_{\mcT_i}$ is the tree approximation covariance matrix for $\bDelta_{i-1}$, i.e. $\bSigma_{\mcT_i} = \text{chowliu}(\bDelta_{i-1})$. 
As shown in figures \ref{fig:t4}, the $i$-stage cascade tree decomposition, transforms the zero-mean Gaussian random vector $\uZ_i$ to the zero-mean Gaussian random vector $\uX$.

\noindent {\bf Remark:} For all $i\geq 1, \; \textnormal{tr}\{\bDelta_i\} = n$. Trace of the CAM, $\bDelta_i$ is equal to $n$, since the covariance matrix $\bSigma_{\mcT_i}$ at each iteration is obtained by the Chow-Liu algorithm and thus satisfies the covariance selection rules \cite{dempster}, i.e. 
$\textnormal{tr}\{(\bDelta_{i-1} - \bSigma_{\mcT_i})\bSigma_{\mcT_i}^{-1}\} = 0$ and thus $\textnormal{tr}\{\bDelta_{i-1}\bSigma_{\mcT_i}^{-1}\} = n$.

In figure \ref{fig:t5} we use the cascade tree decompositions to construct the approximation model.  If we just use one tree, $\bC_{\mcT_1}$ we have a tree approximation. For a cascade of $l$ trees the approximation model is constructed using a backwards algorithm via the following cascade of linear tree approximations;
$(\bC_{\mcT_l} \bC_{\mcT_{l-1}} \ldots \bC_{\mcT_1}) (\bC_{\mcT_l} \bC_{\mcT_{l-1}} \ldots \bC_{\mcT_1})^T= \bSigma_{\mcM_l}$.  We also have following properties in lemma \ref{lem:lem1} and lemma \ref{lem:lem2}.

\begin{Lemma}{Let $\uW \sim \mcN(\uzm , \bI)$, then}
	\label{lem:lem1}
	\begin{itemize}
		\item[(a)] $\mcD(f_{\uZ_i}(\uz) || f_{\uW}(\uw)) = \mcD(f_{\uZ_{i-1}}(\uz) || f_{\uX_{\mcT_i}}(\ux) ),$
		\item[(b)] $\mcD(f_{\uZ_{i-1}}(\uz) || f_{\uX_{\mcT_i}}(\ux) ) \leq \mcD(f_{\uZ_{i-1}}(\uz) || f_{\uW}(\uw) ),$
	\end{itemize}
	where in (b) equality happens when $f_{\uZ_{i-1}}(\uz) = f_{\uW}(\uw)$, i.e $\bDelta_{i-1} = \bI$.
	\proof Proof of part (a) is based on the definition of KL divergence for jointly Gaussian distribution and $\textnormal{tr}\{\bDelta_i\} = n$ as follow
	\begin{align*}
	\mcD(f_{\uZ_{i-1}}(\uz) || f_{\uX_{\mcT_i}}(\ux) ) & = -\frac{1}{2} \textnormal{log} (|\bDelta_{i-1}\bSigma_{\mcT_i}^{-1}|)\\ & = -\frac{1}{2} \textnormal{log} (|\bC_{\mcT_i}^{-1} \bDelta_{i-1} \bC_{\mcT_i}^{-T}|)\\ &  = -\frac{1}{2} \textnormal{log} (|\bDelta_i|)
	\\ & = \mcD(f_{\uZ_i}(\uz) || f_{\uW}(\uw)).
	\end{align*}
	
	 Proof of part (b) follows from the KL divergence definition for Gaussian distributions and
	\begin{align*}
	\mcD(f_{\uZ_{i-1}}(\uz) || f_{\uW}(\uw)) & = \mcD(f_{\uZ_{i-1}}(\uz) || f_{\uX_{\mcT_i}}(\ux) ) \\ & + \mcD(f_{\uX_{\mcT_i}}(\ux) || f_{\uW}(\uw)) \\ & \geq \mcD(f_{\uZ_{i-1}}(\uz) || f_{\uX_{\mcT_i}}(\ux) ).
	\end{align*}
	Equality only happens if $|\bSigma_{\mcT_i}| = 1$. Since $\textnormal{tr}\{\bDelta_{i-1}\} = n$, then the covariance selection rule \cite{dempster} dictates $\textnormal{tr}\{\bSigma_{\mcT_i}\} = n$, and thus the equality only happens if $\bDelta_{i-1} = \bI$.	 \hfill \IEEEQED
\end{Lemma}

Lemma \ref{lem:lem1} states that the distribution of the $i$-th step residue random vector $\uZ_i$  converges to the normal Gaussian random vector $\uW$.
Thus, in the cascade tree model approximation algorithm, we fix the number of cascade stages, $l$, and input the normal Gaussian random vector $\uW$ to the cascade trees with $l$ stages to do model approximation. The $l$-th step model covariance matrix approximation is
$$\bSigma_{\mcM_l} = \bC_{\mcM_l} \, \bC_{\mcM_l}^T$$
where $\bC_{\mcM_l} = \bC_{\mcT_1} \bC_{\mcT_{2}} ... \bC_{\mcT_i} ... \bC_{\mcT_l}$ is the model transformation. Note that, this is a backward construction (figure \ref{fig:t4}).

\begin{Lemma}{KL divergence upper bound.}
	\label{lem:lem2}
	$$\mcD(f_{\uX}(\ux) || f_{\uX_{\mcM_i}}(\ux)) \leq \mcD(f_{\uX}(\ux) || f_{\uX_{\mcM_{i-1}}}(\ux) )$$
	with equality only happens if $\bDelta_{i-1} = \bI$.
	\proof
	\begin{align*}
		\mcD(f_{\uX}(\ux) || f_{\uX_{\mcM_i}}(\ux)) & \stackrel{(a)}{=} \mcD(f_{\uZ_i}(\uz) || f_{\uW}(\uw))\\ & \stackrel{(b)}{=} \mcD(f_{\uZ_{i-1}}(\uz) || f_{\uX_{\mcT_i}}(\ux) ) \\ & \stackrel{(c)}{\leq} \mcD(f_{\uZ_{i-1}}(\uz) || f_{\uW}(\uw) )\\ & \stackrel{(d)}{=} \mcD(f_{\uX}(\ux) || f_{\uX_{\mcM_{i-1}}}(\ux) )
	\end{align*}
	where (a) and (d) are because of the invariance of KL divergence between Gaussian distributions to the transformation; (b) and (c) follow from lemma \ref{lem:lem1}. Equality in (c) holds if $f_{\uZ_{i-1}}(\uz) = f_{\uW}(\uw)$.  \hfill \IEEEQED
	\vspace{0.3cm}
\end{Lemma}

\begin{Theorem}{\bf The Cascade tree decomposition transformation.}
	\label{thm:CTD}
	As the number of cascade trees, i, increases, the KL divergence between the distribution of $\uX$ and the model distribution decreases, i.e. $\mcD(f_{\uX}(\ux) || f_{\uX_{\mcM_i}}(\ux))$ 
	converges to a finite value.
%	goes to zero as $i \rightarrow \infty$.
	\proof Proof follows directly from lemma \ref{lem:lem2} and positivity of KL divergence. \hfill \IEEEQED
\end{Theorem}

\begin{Conjecture}
The KL divergence converges to 0.
\end{Conjecture}
	
\noindent {\bf Remark:} Theorem \ref{thm:CTD} states that the KL divergence between the model approximation and the original distribution decreases as we add more stages of the cascade trees (Lemma \ref{lem:lem2}) and we conjecture it will go to zero as the number of cascade trees goes to infinity. Note that, it is exactly equal to zero if at some iteration of the cascade tree $\bDelta_{i-1} = \bI$.

\section{Algorithm}
\label{sec:alg}

In order to use the cascade tree transformation decomposition framework presented in section \ref{sec:cascade}, we need to pick a factorization scheme. Here we pick the Cholesky factorization. The main reason is that this scheme can preserve the sparsity pattern of the covariance matrix and thus is suitable to run message passing algorithms over factor graphs.
Also, using the Cholesky decomposition, without loss of generality the diagonal coefficients of the symmetric CAM at all cascade stages are equal to one.
Figure \ref{fig:FG} shows a sample tree structured graph and its factor graph representation using the coefficients of the inverse of the Cholesky decomposition matrix, $\bQ$.

\begin{figure*}
	\begin{minipage}{0.4\linewidth}
		\centering
		\resizebox{0.7\linewidth}{0.7\linewidth}{
		\begin{tikzpicture}[node distance = 9em, align = flush center, font = , on grid=false]	
		\tikzstyle{factor} = [ rectangle, draw, fill=green!30, text centered, minimum height=.5em ]
		\tikzstyle{state} = [ circle, draw, fill=white!20, text centered, minimum height=.3em ]
		\tikzstyle{fact} =   [ rounded rectangle, text centered, minimum height=.5em ]	
		
		\node[state] (F1)  {$X_1^1$};
		\node[state, below left	of=F1] (F2) {$X_1^2$};
		\node[state, below right of=F2] (F3)  {$X_1^3$};  
		\node[state, below right of=F1] 		    (F4)  {$X_1^4$};
		\node[state,  below  of=F4]    (F5) {$X_1^5$}; 
		
		\draw	(F1) -- (F2) node[midway, font=, sloped, below] {$\rho{12}$};
		\draw 	(F1) -- (F3) node[midway, font=, sloped, above] {$\rho_{13}$};
		\draw	(F1) -- (F4) node[midway, font=, sloped, above] {$\rho_{14}$};
		\draw	(F4) -- (F5) node[midway, font=, sloped, above] {$\rho_{45}$};
		\end{tikzpicture}
		}		
	\end{minipage}
	\hspace{.05\linewidth}
	\begin{minipage}{0.5\linewidth}
		\centering
		\resizebox{0.7\linewidth}{0.75\linewidth}{
		\begin{tikzpicture} 
		\GraphInit[vstyle=Classic]
		\SetUpVertex[FillColor=white]  
		\Vertex[           LabelOut=false]{$Z_1^1$}
		\Vertex[x=0, y=-2, LabelOut=false]{$Z_1^2$}
		\Vertex[x=0, y=-4, LabelOut=false]{$Z_1^3$}
		\Vertex[x=0, y=-6, LabelOut=false]{$Z_1^4$}
		\Vertex[x=0, y=-8, LabelOut=false]{$Z_1^5$}
		
		\Vertex[x=6, y= 0, LabelOut=false]{$X_1^1$}
		\Vertex[x=6, y=-2, LabelOut=false]{$X_1^2$}
		\Vertex[x=6, y=-4, LabelOut=false]{$X_1^3$}
		\Vertex[x=6, y=-6, LabelOut=false]{$X_1^4$}
		\Vertex[x=6, y=-8, LabelOut=false]{$X_1^5$}
		
		%vars
		\tikzset{VertexStyle/.append style={rectangle}}
		\Vertex[x=2,y= 0,LabelOut=false]{$q_{11}$}
		\Vertex[x=2,y=-2,LabelOut=false]{$q_{22}$}  
		\Vertex[x=2,y=-4,LabelOut=false]{$q_{33}$}  
		\Vertex[x=2,y=-6,LabelOut=false]{$q_{44}$}  
		\Vertex[x=2,y=-8,LabelOut=false]{$q_{55}$}  
		%sums
		\Vertex[x=4,y=-2,LabelOut=false]{$+$}  
		\Vertex[x=4,y=-4,LabelOut=false]{$+$}  
		\Vertex[x=4,y=-6,LabelOut=false]{$+$}  
		\Vertex[x=4,y=-8,LabelOut=false]{$+$}  
		%connections
		\Vertex[x=4,y=-1,LabelOut=false]{$q_{21}$}  
		\Vertex[x=4,y=-3,LabelOut=false]{$q_{31}$}  
		\Vertex[x=4,y=-5,LabelOut=false]{$q_{41}$}  
		\Vertex[x=4,y=-7,LabelOut=false]{$q_{54}$}
		
		\path [line] (0.4, 0) -- (1.7, 0);
		\path [line] (0.4,-2) -- (1.7,-2);
		\path [line] (0.4,-4) -- (1.7,-4);
		\path [line] (0.4,-6) -- (1.7,-6);
		\path [line] (0.4,-8) -- (1.7,-8);
		
		\path [line] (2.3,0) -- (5.6,0);
		
		\path [line] (2.3,-2) -- (3.7,-2);
		\path [line] (2.3,-4) -- (3.7,-4);
		\path [line] (2.3,-6) -- (3.7,-6);
		\path [line] (2.3,-8) -- (3.7,-8);
		
		\path [line] (4.3,-2) -- (5.7,-2);
		\path [line] (4.3,-4) -- (5.7,-4);
		\path [line] (4.3,-6) -- (5.7,-6);
		\path [line] (4.3,-8) -- (5.7,-8);
		
		\path [line, dashed] (4,-1.3) -- (4,-1.7);
		\path [line, dashed] (4,-3.3) -- (4,-3.7);
		\path [line, dashed] (4,-5.3) -- (4,-5.7);
		\path [line, dashed] (4,-7.3) -- (4,-7.7);

		\draw [line, dashed] (6, -0.4) .. controls (4.5,-1) and (5,-1) .. (4.3,-1);
		\draw [line, dashed] (6, -0.4) .. controls (4.5,-1) and (5,-3) .. (4.3,-3);
		\draw [line, dashed] (6, -0.4) .. controls (5,-1) and (5,-5)  .. (4.3,-5);
		\draw [line, dashed] (6, -6.4) .. controls (4.5,-7) and (5,-7) .. (4.3,-7);
		
		%\path [line, dashed] (6, -0.4) -- (6,-1);
		%\path [line, dashed] (6, -1)   -- (4.3,-1);
		\end{tikzpicture}
		}
	\end{minipage}
	\caption{{\bf Left:} Tree representation of the random vector $\uX$ ($\rho_{ij}$'s are the correlation coefficients). {\bf Right:} Factor graph representation with 5 nodes where $\bQ = \bL^{-1}$ and $q_{ij}$'s are the coefficients of the matrix $\bQ$.}
	\label{fig:FG}
\end{figure*}

\tikzstyle{dist}=[ text width=3 em, minimum height=10 mm,text centered,node distance=3 em]
\tikzstyle{block}=[draw,rectangle,  text width=2.5 em, minimum height=10 mm,text centered,node distance=3 em]
\tikzstyle{line}=[draw, -stealth,thick]
\begin{subfigures}
	\begin{figure*}
		\centering	
		\begin{tikzpicture}
		\node [dist] (s) {$\uZ_i$};
		\node [block, right of = s, xshift=2em] (b1) {$\bC_{\mcT_i}$};
		\node [dist, right of = b1, xshift=2em] (b2) {$\uZ_{i-1}$};
		\node [dist, right of = b2, xshift=2em] (M) {\large$\rightarrow$};
		\node [dist, right of = M,  xshift=2em] (s1) {$\uZ_i$};
		\node [block, right of = s1, xshift=2em] (p1) {$\bP_i^{-T}$};		
		\node [block, right of = p1, xshift=2em] (l1) {$\bL_i$};
		\node [block, right of = l1, xshift=2em] (p2) {$\bP_i^{-T}$};
		\node [dist, right of = p2, xshift=2em] (e) {$\uZ_{i-1}$};
		\path [line] (s) -- (b1);
		\path [line] (b1) -- (b2);
		\path [line] (s1) -- (p1);
		\path [line] (p1) -- (l1);
		\path [line] (l1) -- (p2);
		\path [line] (p2) -- (e);
		\end{tikzpicture}
		\caption{{\bf Left:} The $i$-th stage of the model transformation from $\uZ_i$ to $\uZ_{i-1}$. {\bf Right:} The $i$-th stage of the model transformation from $\uZ_i$ to $\uZ_{i-1}$ using proper permutation matrix and the Cholesky decompositions.}
		\label{fig:t6}
		%\end{figure*}
		\vspace{0.5cm}
		%\begin{figure*}
		\centering	
		\begin{tikzpicture}
		\node [dist] (s1) {$\uW$};
		\node [block, right of = s1, xshift=2em] (p1) {$\bP_l^{-T}$};		
		\node [block, right of = p1, xshift=2em] (l1) {$\bL_l$};
		\node [block, right of = l1, xshift=2em] (p2) {$\bP_l^{-T}$};
		\node [dist, right of = p2,  xshift=2em] (M) {...};		
		\node [block, right of = M,  xshift=2em] (p11) {$\bP_1^{-T}$};		
		\node [block, right of = p11, xshift=2em] (l11) {$\bL_1$};
		\node [block, right of = l11, xshift=2em] (p21) {$\bP_1^{-T}$};
		\node [dist, right of = p21, xshift=2em] (e) {$\uX_{\mcM_l}$};
		\path [line] (s1) -- (p1);
		\path [line] (p1) -- (l1);
		\path [line] (l1) -- (p2);
		\path [line] (p2) -- (M);
		\path [line] (M) -- (p11);
		\path [line] (p11) -- (l11);
		\path [line] (l11) -- (p21);
		\path [line] (p21) -- (e);
		\end{tikzpicture}
		\caption{The $l$ stages of cascade tree model approximation using the Cholesky decomposition with proper order (permutation) to keep the sparsity pattern in the inverse of The Cholesky decomposition. The model approximation is generated by passing $\uW \sim \mcN(\uzm, \bI)$ through the $l$ steps cascade trees and is $\uX_{\mcM_l} \sim \mcN(\uzm, \bSigma_{\mcM_l})$.}
		\label{fig:CD_Mapprox}
	\end{figure*}
\end{subfigures}

\subsection{Greedy Model Approximation Algorithm}

Here we present a greedy algorithm based on the cascade trees principle. This algorithm consists of two general steps:
\begin{itemize}
	\item[-] Finding the optimal Chow Liu tree,
	\item[-] Performing the Cholesky decomposition such that it preserves the tree graph structure.
\end{itemize}
Given the symmetric CAM at each iteration of the greedy algorithm, we can efficiently find the optimal tree structure covariance matrix.

\begin{Theorem}
	\label{lem:permu}
	There exists a permutation matrix such that the inverse of the Cholesky decomposition preserves the sparsity pattern (position of zeros) of the inverse of the tree approximation covariance matrix \cite{SparseMatDecomp}. 
	\proof This lemma is a simplified version of the result presented in \cite{SparseMatDecomp} where the Cholesky decomposition preserves the pattern of zeros corresponding to a co-chordal or homogeneous graph associated with a specific type of vertex ordering (permutation matrix).	 \hfill \IEEEQED
\end{Theorem}

Theorem \ref{lem:permu} guarantees the existence of a loop-free factor graph based on the Cholesky decomposition coefficients of the inverse decomposition.

Figure \ref{fig:t6} shows the schematic of the $i$-th stage of the model transformation based on the proper permutation of the Cholesky decomposition. 
To compute matrices $\bP_i$ and $\bL_i$ and reconstruct $\uZ_{i-1}$ from $\uZ_i$, we need to first compute the $i$-th stage tree approximation covariance matrix,  $\bSigma_{\mcT_i} = \textnormal{chowliu}(\bDelta_{i-1})$. 
Next, we use the result of theorem \ref{lem:permu} to find the proper re-order of the nodes. To do that, we look at the graph structure of the tree covariance matrix, $\bSigma_{\mcT_i}$, and pick the nodes such that the graph associated with the subset of the picked nodes is always connected\footnote{If at any step of the algorithm, the tree graph structure associated with $\bSigma_{\mcT_i}$ becomes disconnected, we will seek the same procedure for each of the disjoint segments of the graph.}. After that, we compute the Cholesky decomposition as $L_i = \textnormal{Cholesky}(\bP_i \bSigma_{\mcT_i} \bP_i^T)$ which has a sparse inverse. Next, we permute $L_i$ to get the tree approximation transformation matrix, $\bC_{\mcT_i}$.
This process is shown in figure \ref{fig:CD_Mapprox}. The model approximation covariance matrix after the $l$-th iteration is given as follow
$$\bSigma_{\mcM_l} =\bC_{\mcM_l} \, \bC_{\mcM_l}^T$$
where $\bC_{\mcM_l} = \bP_1^{-1} \bL_1 \bP_1^{-T} \bP_2^{-1} \bL_2 \bP_2^{-T} \ldots \bP_l^{-1} \bL_l \bP_l^{-T}$.

Using the Cholesky decomposition with the proper permutation matrix at each iteration enables us to draw loop-free factor graph at each iteration of the cascade trees' framework. The factor graph representation is useful in order to run message passing algorithm and loopy GBP over the overall loopy factor graph.
The greedy algorithm based on the Cholesky decomposition presented in figure \ref{fig:CD_Mapprox} is as follow:
\\
\hrule
\vspace{0.2cm}
{\bf Greedy Model Approximation Algorithm using Cascade Trees' Framework and the Cholesky Decomposition}
\vspace{0.15cm}
\hrule
\vspace{0.15cm}
\begin{itemize}
	\item{Initialization Step [$i=0$]:} 
	\begin{itemize}
		\item $\bDelta_{0} = \bSigma$
	\end{itemize}
	\vspace{0.1cm}
	\item{Continue updating [$i$-th Step]:} 
	\begin{itemize}
		\item $i \leftarrow i+1$
		\vspace{0.1cm}
		\item $\bSigma_{\mcT_i} = \textnormal{chowliu}(\bDelta_{i-1})$
		\vspace{0.1cm}
		\item Given $\bSigma_{\mcT_i}$, compute the proper node ordering and construct the permutation matrix, $\bP_i$.
		\vspace{0.1cm}
		\item $L_i = \textnormal{Cholesky}(\bP_i \bSigma_{\mcT_i} \bP_i^T)$
		\vspace{0.1cm}
		\item $\bC_{\mcT_i} = \bP_i^{-1} \bL_i \bP_i^{-T}$
		\vspace{0.1cm}
%		\item $\bC_{\mcM_i} = \bC_{\mcM_{i-1}} \bC_{\mcT_i}$
 		\item $\bDelta_i = \bC_{\mcT_i}^{-1} \bDelta_{i-1} \bC_{\mcT_i}^{-T}$
		\vspace{0.1cm}
	\end{itemize}
	\item Stopping criterion\footnote{We can stop the algorithm sooner if for some $i$ the KL divergence goal is satisfied, i.e. $\mcD(f_{\uZ_i}(\uz) || f_{\uW}(\uw)) \leq \delta$ where $\delta$ is the maximum KL divergence between the original distribution and the approximated model distribution.}: $i \leq l$
	\vspace{0.15cm}
	\item Output $\bL_i$'s and $\bP_i$'s as well as the approximated model covariance matrix $\bSigma_{\mcM_l} = \bC_{\mcM_l}\bC_{\mcM_l}^T$ where $\bC_{\mcM_l} = \bC_{\mcT_1} \ldots \bC_{\mcT_l}$ for some $i$ satisfying the stopping criterion.
	\vspace{0.1cm}
\end{itemize}
\hrule	
\vspace{0.5cm}

\noindent {\bf Remark:} In any step of the algorithm, if the graph correspond to the $\bDelta_i$ become disconnected, we will do tree approximation for each of the connected subgraphs.

\begin{Theorem}
\label{thm:thm5}
	There exists a cascade tree approximation algorithm to generate the model approximation $\mcM'_i$ such that after at most $n-1$ iteration, the model approximation error (KL divergence) is exactly equal to zero, i.e. $\bSigma = \bSigma_{\mcM_{n-1}}$. In other words,
	$$\mcD(f_{\uX}(\ux) || f_{\uX_{\mcM'_{n-1}}}(\ux)) = 0.$$
	\proof Proof is given in appendix \ref{apx:apx1}.	
	\hfill \IEEEQED
\end{Theorem}

\begin{Theorem}{\bf Diagonal Coefficients of the Symmetric CAM.}
	\label{lem:diag1}
	Diagonal coefficients of the symmetric CAM at each step of the the greedy model approximation algorithm using the Cholesky factorization, are equal to one. 
	\proof Proof is given in appendix \ref{apx:apx2}.	 \hfill \IEEEQED
\end{Theorem}

Theorem \ref{lem:diag1} shows that if we pick the Cholesky factorization and we follow the greedy model approximation algorithm presented here, then diagonal coefficients of the approximated matrix is always the same as diagonal coefficients of the covariance matrix, $\bSigma$, i.e. the proposed algorithm preserves variances.

\subsection{Complexity of the cascade tree algorithm}

Chow-Liu algorithm has the complexity of $\mathcal{O}(n^2 \log(n))$ while the Cholesky decomposition complexity is $\mathcal{O}(n^3)$. Thus, the overall complexity of the algorithm is $\mathcal{O}(l \, n^3)$ since we run the algorithm for at most $l$ cascade stages. Moreover, the maximum number of edges in the resulting factor graph is $l n$.

\subsection{Example with 5 nodes}

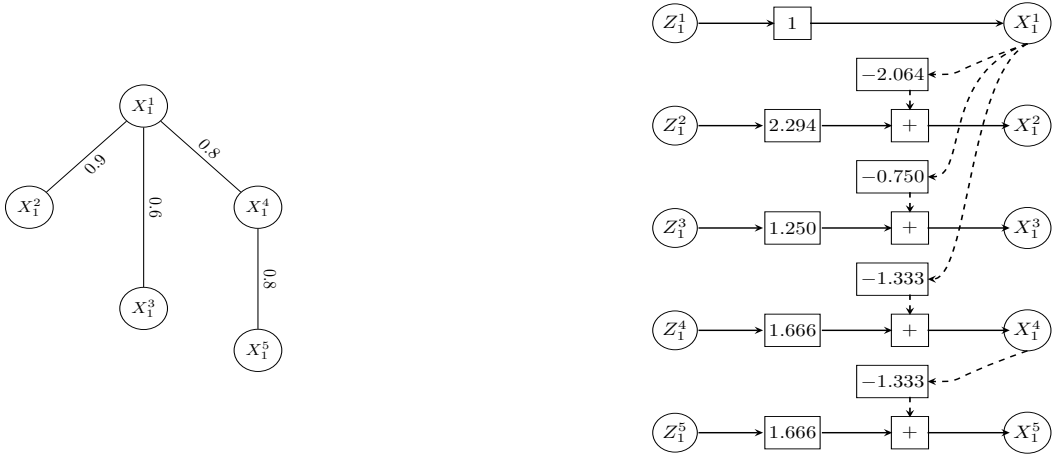
\begin{figure*}[t]
	\begin{minipage}{0.3\linewidth}
		\centering
		\resizebox{0.7\linewidth}{0.7\linewidth}{
		\begin{tikzpicture}[node distance = 9em, align = flush center, font = , on grid=false]	
		\tikzstyle{factor} = [ rectangle, draw, fill=green!30, text centered, minimum height=.5em ]
		\tikzstyle{state} = [ circle, draw, fill=white!20, text centered, minimum height=.3em ]
		\tikzstyle{fact} =   [ rounded rectangle, text centered, minimum height=.5em ]	
		
		\node[state] (F1)  {$X_1^1$};
		\node[state, below left  of=F1] (F2)  {$X_1^2$};
		\node[state, below right of=F2] (F3)  {$X_1^3$};  
		\node[state, below right of=F1] (F4)  {$X_1^4$};
		\node[state, below       of=F4] (F5)  {$X_1^5$}; 
		
		\draw	(F1) -- (F2) node[midway, font=, sloped, below] {$0.9$};
		\draw 	(F1) -- (F3) node[midway, font=, sloped, above] {$0.6$};
		\draw	(F1) -- (F4) node[midway, font=, sloped, above] {$0.8$};
		\draw	(F4) -- (F5) node[midway, font=, sloped, above] {$0.8$};
		\end{tikzpicture}	
		}	
	\end{minipage}
	\hspace{.05\linewidth}
	\begin{minipage}{0.6\linewidth}
		\centering
		\resizebox{0.5\linewidth}{0.55\linewidth}{
		\begin{tikzpicture} 
		\GraphInit[vstyle=Classic]
		\SetUpVertex[FillColor=white]  
		\Vertex[           LabelOut=false]{$Z_1^1$}
		\Vertex[x=0, y=-2, LabelOut=false]{$Z_1^2$}
		\Vertex[x=0, y=-4, LabelOut=false]{$Z_1^3$}
		\Vertex[x=0, y=-6, LabelOut=false]{$Z_1^4$}
		\Vertex[x=0, y=-8, LabelOut=false]{$Z_1^5$}
		
		\Vertex[x=6, y= 0, LabelOut=false]{$X_1^1$}
		\Vertex[x=6, y=-2, LabelOut=false]{$X_1^2$}
		\Vertex[x=6, y=-4, LabelOut=false]{$X_1^3$}
		\Vertex[x=6, y=-6, LabelOut=false]{$X_1^4$}
		\Vertex[x=6, y=-8, LabelOut=false]{$X_1^5$}
		
		%vars
		\tikzset{VertexStyle/.append style={rectangle}}
		\Vertex[x=2,y= 0,LabelOut=false]{$1$}
		\Vertex[x=2,y=-2,LabelOut=false]{$2.294$}  
		\Vertex[x=2,y=-4,LabelOut=false]{$1.250$}  
		\Vertex[x=2,y=-6,LabelOut=false]{$1.666$}  
		\Vertex[x=2,y=-8,LabelOut=false]{$1.666$}  
		%sums
		\Vertex[x=4,y=-2,LabelOut=false]{$+$}  
		\Vertex[x=4,y=-4,LabelOut=false]{$+$}  
		\Vertex[x=4,y=-6,LabelOut=false]{$+$}  
		\Vertex[x=4,y=-8,LabelOut=false]{$+$}  
		%connections
		\Vertex[x=3.7,y=-1,LabelOut=false]{$-2.064$}  
		\Vertex[x=3.7,y=-3,LabelOut=false]{$-0.750$}  
		\Vertex[x=3.7,y=-5,LabelOut=false]{$-1.333$}  
		\Vertex[x=3.7,y=-7,LabelOut=false]{$-1.333$}
		
		\path [line] (0.4, 0) -- (1.7, 0);
		\path [line] (0.4,-2) -- (1.5,-2);
		\path [line] (0.4,-4) -- (1.5,-4);
		\path [line] (0.4,-6) -- (1.5,-6);
		\path [line] (0.4,-8) -- (1.5,-8);
		
		\path [line] (2.3,0) -- (5.6,0);
		
		\path [line] (2.5,-2) -- (3.7,-2);
		\path [line] (2.5,-4) -- (3.7,-4);
		\path [line] (2.5,-6) -- (3.7,-6);
		\path [line] (2.5,-8) -- (3.7,-8);
		
		\path [line] (4.3,-2) -- (5.7,-2);
		\path [line] (4.3,-4) -- (5.7,-4);
		\path [line] (4.3,-6) -- (5.7,-6);
		\path [line] (4.3,-8) -- (5.7,-8);
		
		\path [line, dashed] (4,-1.3) -- (4,-1.7);
		\path [line, dashed] (4,-3.3) -- (4,-3.7);
		\path [line, dashed] (4,-5.3) -- (4,-5.7);
		\path [line, dashed] (4,-7.3) -- (4,-7.7);

		\draw [line, dashed] (6, -0.4) .. controls (4.5,-1) and (5,-1) .. (4.3,-1);
		\draw [line, dashed] (6, -0.4) .. controls (4.5,-1) and (5,-3) .. (4.3,-3);
		\draw [line, dashed] (6, -0.4) .. controls (5,-1) and (5,-5)  .. (4.3,-5);
		\draw [line, dashed] (6, -6.4) .. controls (4.5,-7) and (5,-7) .. (4.3,-7);
		
		%\path [line, dashed] (6, -0.4) -- (6,-1);
		%\path [line, dashed] (6, -1)   -- (4.3,-1);
		\end{tikzpicture}
	}
	\end{minipage}
	\caption{First stage cascade tree representation for the 5 nodes example and its Factor graph representation.}
	\label{fig:FG_S1}
\end{figure*}

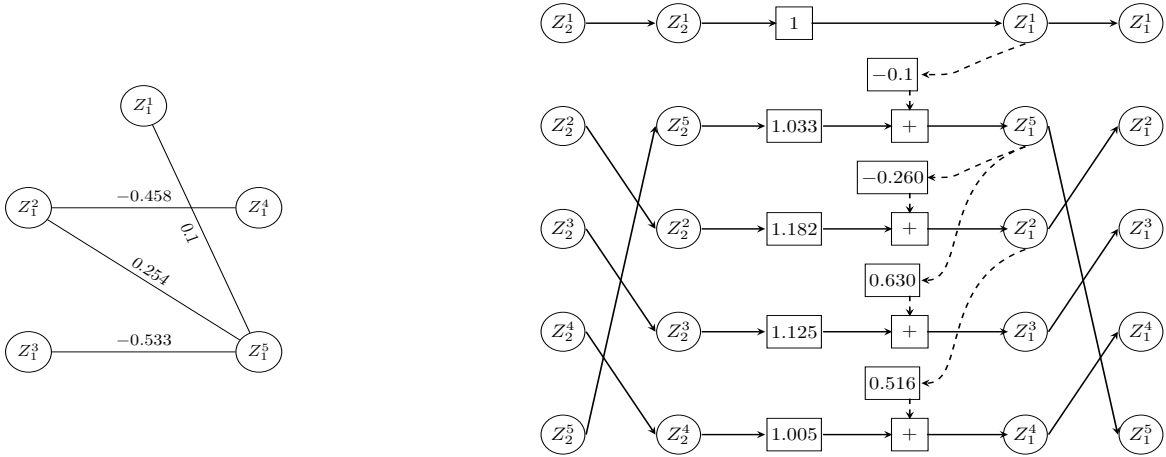
\begin{figure*}
	\begin{minipage}{0.3\linewidth}
		\centering
		\resizebox{0.7\linewidth}{0.7\linewidth}{
		\begin{tikzpicture}[node distance = 9em, align = flush center, font = , on grid=false]	
		\tikzstyle{factor} = [ rectangle, draw, fill=green!30, text centered, minimum height=.5em ]
		\tikzstyle{state} = [ circle, draw, fill=white!20, text centered, minimum height=.3em ]
		\tikzstyle{fact} =   [ rounded rectangle, text centered, minimum height=.5em ]	
		
		\node[state] (F1)  {$Z_1^1$};
		\node[state, below left	of=F1] (F2) {$Z_1^2$};
		\node[state, below  of=F2] (F3)  {$Z_1^3$};  
		\node[state, below right of=F1]  (F4) {$Z_1^4$};
		\node[state,  below  of=F4]    (F5) {$Z_1^5$}; 
		
		\draw	(F1) -- (F5) node[midway, font=, sloped, below] {$0.1$};
		\draw 	(F5) -- (F2) node[midway, font=, sloped, above] {$0.254$};
		\draw	(F5) -- (F3) node[midway, font=, sloped, above] {$-0.533$};
		\draw	(F2) -- (F4) node[midway, font=, sloped, above] {$-0.458$};
		\end{tikzpicture}	
	}	
	\end{minipage}
	\hspace{.05\linewidth}
	\begin{minipage}{0.6\linewidth}
		\centering
		\resizebox{0.77\linewidth}{0.55\linewidth}{
		\begin{tikzpicture} 
		\GraphInit[vstyle=Classic]
		\SetUpVertex[FillColor=white]  
		\Vertex[x=-2, y= 0, LabelOut=false]{$Z_2^1$}
		\Vertex[x=-2, y=-2, LabelOut=false]{$Z_2^2$}
		\Vertex[x=-2, y=-4, LabelOut=false]{$Z_2^3$}
		\Vertex[x=-2, y=-6, LabelOut=false]{$Z_2^4$}
		\Vertex[x=-2, y=-8, LabelOut=false]{$Z_2^5$}
		
		\Vertex[           LabelOut=false]{$Z_2^1$}
		\Vertex[x=0, y=-2, LabelOut=false]{$Z_2^5$}
		\Vertex[x=0, y=-4, LabelOut=false]{$Z_2^2$}
		\Vertex[x=0, y=-6, LabelOut=false]{$Z_2^3$}
		\Vertex[x=0, y=-8, LabelOut=false]{$Z_2^4$}
		
		\Vertex[x=6, y= 0, LabelOut=false]{$Z_1^1$}
		\Vertex[x=6, y=-2, LabelOut=false]{$Z_1^5$}
		\Vertex[x=6, y=-4, LabelOut=false]{$Z_1^2$}
		\Vertex[x=6, y=-6, LabelOut=false]{$Z_1^3$}
		\Vertex[x=6, y=-8, LabelOut=false]{$Z_1^4$}
		
		\Vertex[x=8, y= 0, LabelOut=false]{$Z_1^1$}
		\Vertex[x=8, y=-2, LabelOut=false]{$Z_1^2$}
		\Vertex[x=8, y=-4, LabelOut=false]{$Z_1^3$}
		\Vertex[x=8, y=-6, LabelOut=false]{$Z_1^4$}
		\Vertex[x=8, y=-8, LabelOut=false]{$Z_1^5$}
				
		%vars
		\tikzset{VertexStyle/.append style={rectangle}}
		\Vertex[x=2,y= 0,LabelOut=false]{$1$}
		\Vertex[x=2,y=-2,LabelOut=false]{$1.033$}  
		\Vertex[x=2,y=-4,LabelOut=false]{$1.182$}  
		\Vertex[x=2,y=-6,LabelOut=false]{$1.125$}  
		\Vertex[x=2,y=-8,LabelOut=false]{$1.005$}  
		%sums
		\Vertex[x=4,y=-2,LabelOut=false]{$+$}  
		\Vertex[x=4,y=-4,LabelOut=false]{$+$}  
		\Vertex[x=4,y=-6,LabelOut=false]{$+$}  
		\Vertex[x=4,y=-8,LabelOut=false]{$+$}  
		%connections
		\Vertex[x=3.7,y=-1,LabelOut=false]{$-0.1$}  
		\Vertex[x=3.7,y=-3,LabelOut=false]{$-0.260$}  
		\Vertex[x=3.7,y=-5,LabelOut=false]{$0.630$}  
		\Vertex[x=3.7,y=-7,LabelOut=false]{$0.516$}
		
		\path [line] (0.4, 0) -- (1.7, 0);
		\path [line] (0.4,-2) -- (1.5,-2);
		\path [line] (0.4,-4) -- (1.5,-4);
		\path [line] (0.4,-6) -- (1.5,-6);
		\path [line] (0.4,-8) -- (1.5,-8);
		
		\path [line] (2.3,0) -- (5.6,0);
		
		\path [line] (2.5,-2) -- (3.7,-2);
		\path [line] (2.5,-4) -- (3.7,-4);
		\path [line] (2.5,-6) -- (3.7,-6);
		\path [line] (2.5,-8) -- (3.7,-8);
		
		\path [line] (4.3,-2) -- (5.7,-2);
		\path [line] (4.3,-4) -- (5.7,-4);
		\path [line] (4.3,-6) -- (5.7,-6);
		\path [line] (4.3,-8) -- (5.7,-8);
		
		\path [line, dashed] (4,-1.3) -- (4,-1.7);
		\path [line, dashed] (4,-3.3) -- (4,-3.7);
		\path [line, dashed] (4,-5.3) -- (4,-5.7);
		\path [line, dashed] (4,-7.3) -- (4,-7.7);

		\draw [line, dashed] (6, -0.4) .. controls (4.5,-1) and (5,-1) .. (4.2,-1);
		\draw [line, dashed] (6, -2.4) .. controls (4.5,-3) and (5,-5) .. (4.2,-5);
		\draw [line, dashed] (6, -2.4) .. controls (4.5,-3) and (5,-3) .. (4.3,-3);
		\draw [line, dashed] (6, -4.4) .. controls (4.5,-5) and (5,-7) .. (4.2,-7);
		
		\path [line] (6.4, 0) -- (7.6, 0);
		\path [line] (6.4,-2) -- (7.6,-8);
		\path [line] (6.4,-4) -- (7.6,-2);
		\path [line] (6.4,-6) -- (7.6,-4);
		\path [line] (6.4,-8) -- (7.6,-6);	
		
		\path [line] (-1.6, 0) -- (-0.4, 0);
		\path [line] (-1.6,-2) -- (-0.4,-4);
		\path [line] (-1.6,-4) -- (-0.4,-6);
		\path [line] (-1.6,-6) -- (-0.4,-8);
		\path [line] (-1.6,-8) -- (-0.4,-2);	
					
		%\path [line, dashed] (6, -0.4) -- (6,-1);
		%\path [line, dashed] (6, -1)   -- (4.3,-1);
		\end{tikzpicture}
	}
	\end{minipage}
	\caption{Second stage cascade tree representation for the 5 nodes example and its Factor graph representation.}
	\label{fig:FG_S2}
\end{figure*}

In this example, we start with a zero mean Gaussian distribution with covariance matrix $\bSigma$ (or $\bDelta_0$) for random vector $\uX$ as follows
\begin{equation*}
\bSigma = 
\begin{bmatrix}
1   &   0.9   &   0.6   &   0.8  &  0.7 \\
0.9 &    1    &   0.5   &   0.6  &  0.6 \\
0.6 &   0.5   &    1    &   0.4  &  0.1 \\
0.8 &   0.6   &   0.4   &    1   &  0.8 \\
0.7 &   0.6   &   0.1   &   0.8  &   1  
\end{bmatrix}
.
\end{equation*}
We want to approximate the random vector $\uX$ with $2$ stages of cascade trees as $\uX_\mcM$. First step tree approximation covariance matrix $\bT_1$ is 
\begin{equation*}
\bT_1 = 
\begin{bmatrix}
1   &   0.9   &   0.6   &   0.8  &  0.64 \\
0.9 &    1    &   0.54  &   0.72 &  0.576\\
0.6 &   0.54  &    1    &   0.48 &  0.374\\
0.8 &   0.72  &   0.48  &    1   &  0.8  \\
0.64&  0.576  &  0.374  &   0.8  &   1  
\end{bmatrix}
\end{equation*}
while its Cholesky decomposition inverse, $\bQ_1$ is
\begin{equation*}
\bQ_1 = 
\begin{bmatrix}
1   &    0    &    0    &    0   &   0  \\
-2.064 &  2.294  &    0    &    0   &   0  \\
-0.75  &     0   &  1.25   &    0   &   0  \\
-1.333 &     0   &    0    &  1.666 &   0  \\
0   &     0   &    0    & -1.333 & 1.666  
\end{bmatrix}
,
\end{equation*}
and the permutation matrix, $\bP_1$ is identity.
To proceed to the second stage, we first compute the symmetric CAM, $\bDelta_1$ as
\begin{equation*}
\bDelta_1 = 
\begin{bmatrix}
1   &    0    &    0    &    0   &  0.1  \\
0   &    1    & -0.1147 & -0.458 &  0.252\\
0   & -0.114  &    1    & -0.166 &  0.374\\
0   & -0.458  & -0.458  &    1   & -0.133\\
0.1 &  0.252  &  0.252  & -0.133 &   1  
\end{bmatrix}
.
\end{equation*}
The CAM matrix, $\bDelta_1$, is the covariance matrix of the residue random vector, $\uZ_1$, and $\uZ_1 = \bQ_1 \uX$ or equivalently $\uX = \bC_{\mcT_1} \uZ_1$ where $\bC_{\mcT_1} = \bQ_1^{-1}$. Also, the KL divergence for the first step is $\mcD(\uX || \uX_{\mcM_1}) = 0.375$.
Then, the second step tree approximation covariance matrix $\bT_2$ is 
\begin{equation*}
\bT_2 = 
\begin{bmatrix}
1     &  0.025  & -0.053  & -0.011 &  0.1  \\
0.025 &    1    & -0.134  & -0.458 &  0.252\\
-0.053& -0.134  &    1    &  0.061 & -0.533\\
-0.011& -0.458  &   0.061 &    1   & -0.115\\
0.1   &  0.252  &  -0.533 & -0.115 &   1  
\end{bmatrix}
\end{equation*}
while its Cholesky decomposition inverse, $\bQ_2$, which is computed using 
$$\bQ_2 = \bP_2^T \, \textnormal{Cholesky}(\bP_2\bT_2\bP_2^T)^{-1} \bP_2$$
is as follow
\begin{equation*}
\bQ_2 = 
\begin{bmatrix}
1    &    0   &    0    &    0   &   0   \\
0    & 1.033  &    0    &    0   & -0.260\\
0    &    0   &  1.182  &    0   & 0.630 \\
0    & 0.516  &    0    &  1.125 &   0   \\
-0.1 &    0   &    0    &    0   & 1.005  
\end{bmatrix}
,
\end{equation*}
where the permutation matrix, $\bP_2$, is
\begin{equation*}
\bP_2 = 
\begin{bmatrix}
1    &    0   &    0    &    0   &   0   \\
0    &    0   &    0    &    0   &   1   \\
0    &    1   &    0    &    0   &   0   \\
0    &    0   &    1    &    0   &   0   \\
0    &    0   &    0    &    1   &   0  
\end{bmatrix}
.
\end{equation*}
Furthermore, the residue random vectors, $\uZ_1$ and $\uZ_2$, have the following relationship $\uZ_2 = \bQ_2 \uZ_1$ or equivalently $\uZ_1 = \bC_{\mcT_2} \uZ_2$ where $\bC_{\mcT_2} = \bQ_2^{-1}$. To approximate the model random vector $\uX_{\mcM_2}$ using two stage of the cascade tree, we replace the second residue random vector, $\uZ_2$, with the random vector $\uW$. Also, the KL divergence for the second step is $\mcD(\uX || \uX_{\mcM_2}) = 0.051$.
Figure \ref{fig:FG_S1} shows the Chow-Liu tree and the factor graph representation for of it and figure \ref{fig:FG_S2} shows the second stage of the algorithm. Since the permutation matrix is not identity in the second step of the algorithm, we need to change the ordering as it is shown in figure \ref{fig:FG_S2}.
For this example a cascade of two trees  produces a linear transformation that approximates the Gaussian vector $\uX$ closely.

\begin{figure*}[t]
%	\centering
	\hspace{-1.1cm}
	\includegraphics[width=1.15\linewidth]{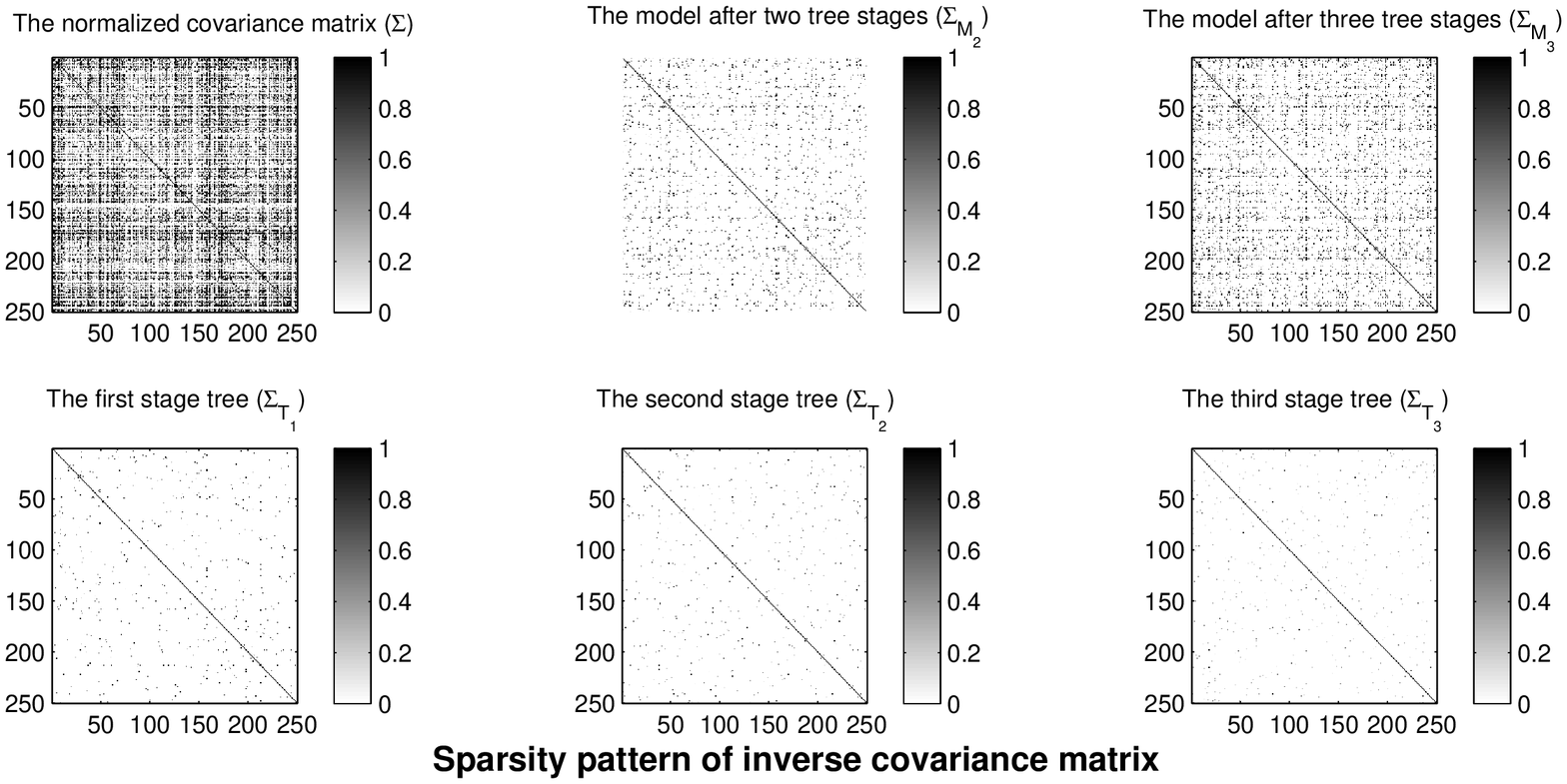}
	\caption{Gray scaled, sparsity pattern for the inverse of a randomly generated, synthetic covariance matrix. {\bf Top left:} Inverse of the original normalized covariance matrix, {\bf Bottom left:} Inverse of the first stage tree approximation and first model. {\bf Top middle:} Inverse of the second approximated model, {\bf Bottom middle:} Inverse of the second stage tree approximation. {\bf Top right:} Inverse of the third approximated model, {\bf Bottom right:} Inverse of the third stage tree approximation.}
	\label{fig:sparsity_pattern_random_sparse_cov_n250}
\end{figure*}

\section{Simulation Results and Discussion}
\label{sec:sim}

In this section, we consider some examples of covariance matrices for Gaussian random vector $\uX$. We present some simulation results on both synthetically generated covariance matrices and the covariance matrix generated from the island of Oahu real solar dataset. We also present the simulation results on the performance of the cascade tree decomposition transformation framework using different factorization methods. Specially, we look at the performance of the presented algorithm in section \ref{sec:alg} which is based on the Cholesky decomposition and the proper permutation to keep the sparsity pattern in the inverse of the Cholesky factorization. We also look at the performance of the singular value decomposition and the Cholesky factorization without the proper permutation (does not keep the sparsity pattern). Looking at other covariance matrices factorization methods gives some insight on how good is the performance of the greedy algorithm which is presented in section \ref{sec:alg}.

\noindent {\bf Remark:} In all of our simulation results we only consider 16 digits precision after the floating point.

% In the simulation section \ref{sec:sim}, we will discuss the performance of other decompositions, although it is not clear how to keep the sparsity pattern in the decomposition.

\subsection{Synthetic data}

We randomly generate synthetic covariance matrix such that its graphical structure has about half of all possible edges and then we normalize it to have ones along the diagonal.

Figure \ref{fig:sparsity_pattern_random_sparse_cov_n250} plots the gray scaled, sparsity pattern for the inverse of a randomly generated, synthetic covariance matrix and various approximations of it. 
The top left plot shows the inverse of the original normalized covariance matrix. The graph associated with this inverse covariance matrix has around $\frac{n^2}{2}$ number of edges. 
The bottom left plot indicates the first stage of the cascade tree approximation (the optimal Chow-Liu solution) or the approximated model after the first stage of the greedy algorithm.
The top middle plot shows the inverse of the second approximated model while the plot on the bottom middle indicates the inverse of the second stage tree approximation. 
The plot on the top right indicates the sparsity pattern of the inverse of third approximated model, while the bottom right plot shows the third stage tree approximation.
In Figure \ref{fig:sparsity_pattern_random_sparse_cov_n250} the  Chow-Liu tree approximation shown in the bottom left is a poor approximation of the top left plot by comparing the two gray-scale plots.  The top middle plot is a better approximation of the top left plot and the top right plot provides the best approximation to the top left plot. The top right plot consists of the cascade of three trees having a maximum of $3\times249$ edges as compared to the top left plot which represents a graph with more than $30000$ edges.

\begin{figure*}[ht]
	\centering
	\includegraphics[width=.9\linewidth]{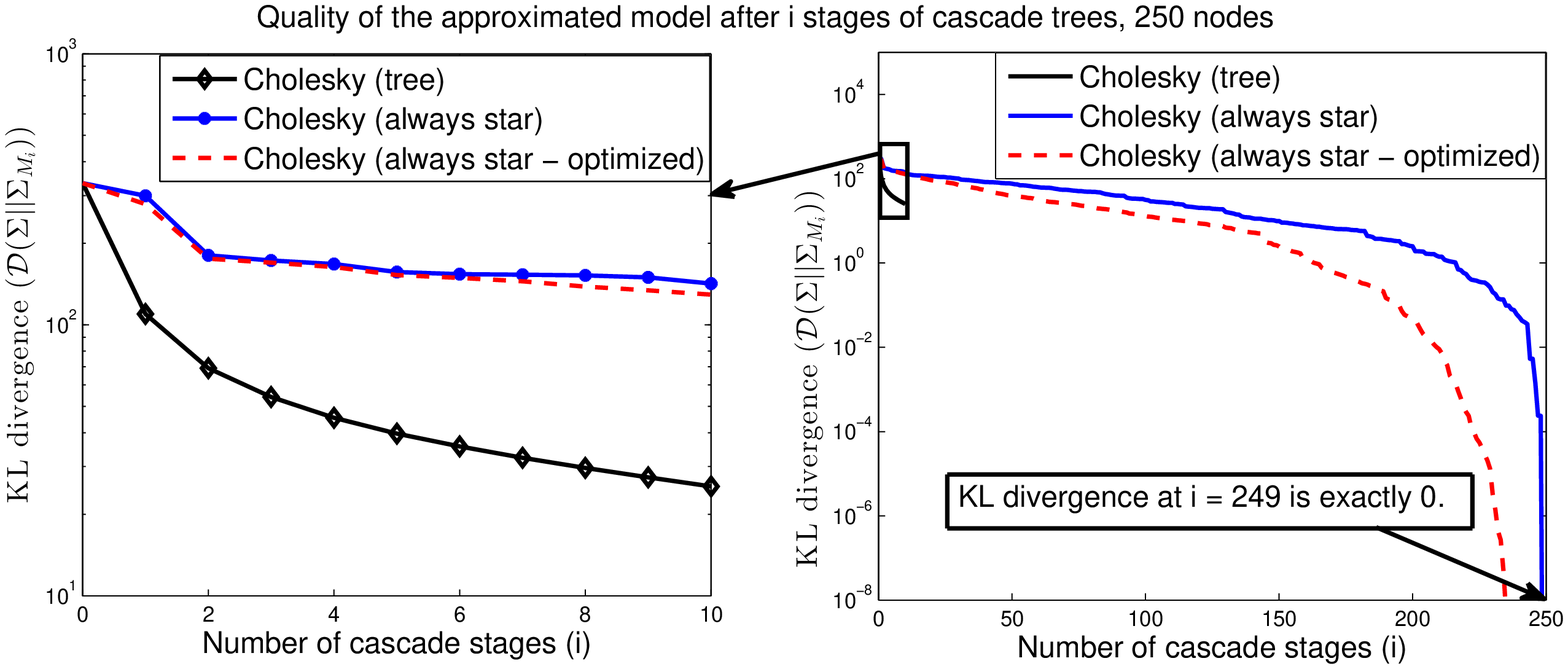}
	\caption{KL divergence between the distribution of the random vector $\uX$ and the model distribution after the $i$-th step of the cascade approximation v.s. the index of the cascade trees, $i$, for a graph with 250 nodes. {\bf Right:} Comparing the performance of three different tree structures, the optimal Chow liu tree, the Star tree without permutation, and the optimal star tree with permutation, as we add more cascade steps. {\bf left:} Zoomed into 10 cascade trees decompositions.}
	\label{fig:KLd_vs_i_random_sparse_cov_cholesky_n250_itr250_zoomed}
\end{figure*}

\begin{figure*}
	\begin{minipage}{0.45\linewidth}
		\centering
		\includegraphics[width=1\linewidth]{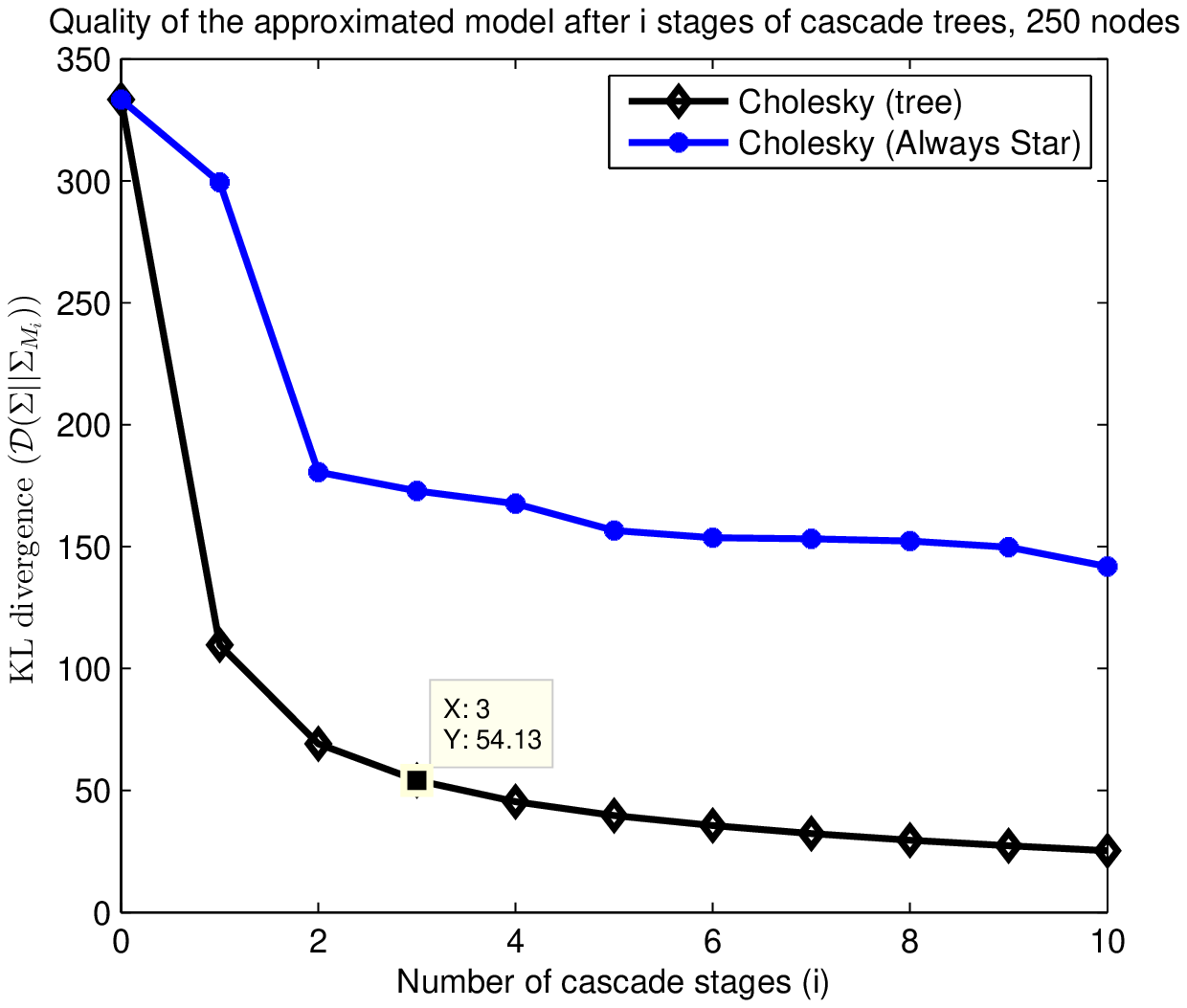}
		\caption{KL divergence between the distribution of the random vector $\uX$ and the model distribution after the $i$-th step of the cascade approximation v.s. the index of the cascade trees, $i$, for a graph with 250 nodes.}
		\label{fig:KLd_vs_i_random_sparse_cov_cholesky_n250_linear}
	\end{minipage}
	\hspace{0.1\linewidth}
	\begin{minipage}{0.45\linewidth}
		\centering
		\includegraphics[width=1\linewidth]{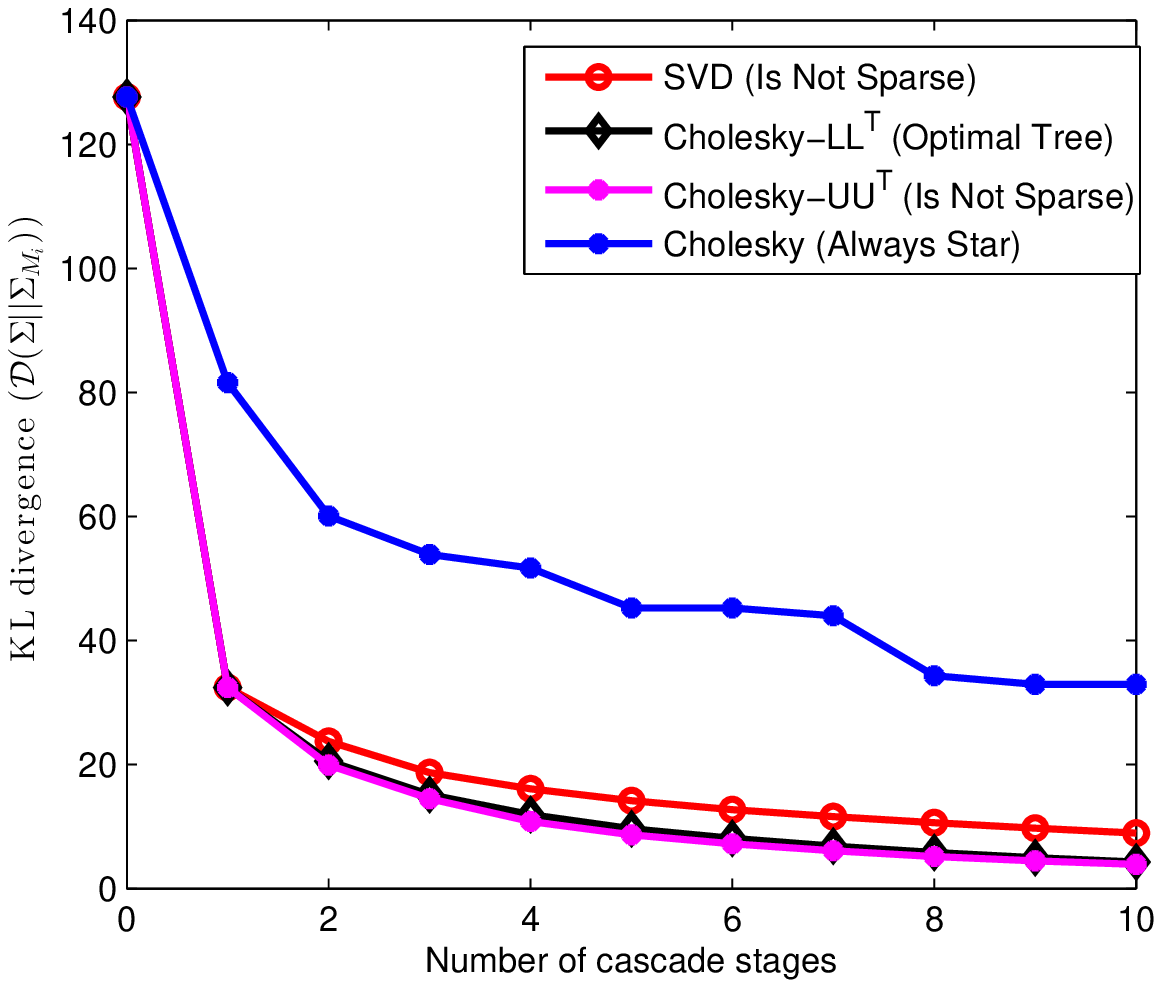}
		\caption{KL divergence between the distribution of the random vector $\uX$ and the model distribution after the $i$-th step of the cascade approximation v.s. the index of the cascade trees, $i$ for a graph with 100 nodes using different decompositions.}
		\label{fig:KLd_vs_i_random_sparse_cov_different_sqrts_n100}
	\end{minipage}
\end{figure*}

Figure \ref{fig:KLd_vs_i_random_sparse_cov_cholesky_n250_itr250_zoomed} plots the log-scaled KL divergence between the random vector $\uX$ and the approximation model vector $\uX_\mcM$ after the $i$-th step of the cascade trees approximation with respect to the number of cascade trees transformation that are used in the approximation, $i$. This figure plots the result of the cascade trees decomposition algorithm for the performance of three different tree structures, the optimal Chow liu tree, the Star tree without permutation, and the optimal star tree with permutation, as we add more cascade steps. The left  plot compares the performance of the cascade trees approximation for different choices of tree structures after 10 steps of the cascade trees decompositions, while the right plot runs the cascade trees algorithm for $249$ steps. 
Looking only at the KL divergence we can easily see that using the greedy algorithm presented in section \ref{sec:alg} clearly has a better performance when we only have small number of cascade stages. On the other hand, running the cascade tree framework using the star tree approximation at each stage for $249$ stages, the KL divergence goes to zero. Note that, figure \ref{fig:KLd_vs_i_random_sparse_cov_cholesky_n250_linear} plots the KL divergence in linear scale. If we compare the Chow-Liu tree to a cascade of two trees/ three trees the KL divergence decreases by respectively 35\%/ 50\% (figure \ref{fig:KLd_vs_i_random_sparse_cov_cholesky_n250_linear}).

\noindent {\bf Remark:} In the $i$-th iteration of the always star approximation we picked node $i$ as the star node to do the approximation without any optimization (permutation matrix is equal to the identity matrix).

Figure \ref{fig:KLd_vs_i_random_sparse_cov_different_sqrts_n100} plots the KL divergence between the random vector $\uX$ and the approximation model vector $\uX_\mcM$ after the $i$-th step of the cascade trees approximation with respect to the number of cascade trees transformation that are used in the approximation, $i$, for a graph of 100 nodes. This figure plots the result of the cascade trees framework with different decompositions such as the Cholesky $\bL \bL^T$ (keep the sparsity), the Cholesky $\bU \bU^T$ (does not keep the sparsity) and the SVD. From figure \ref{fig:KLd_vs_i_random_sparse_cov_different_sqrts_n100} we see that three of the decomposition transformations perform similarly with the star decomposition transformation performing the worse.

%\begin{figure}[ht]
%	\centering
%	\includegraphics[width=.8\linewidth]{KLd_vs_i_random_sparse_cov_cholesky_n250_linear.eps}
%	\caption{KL divergence between the distribution of the random vector $\uX$ and the model distribution after the $i$-th step of the cascade approximation v.s. the index of the cascade trees, $i$, for a graph with 250 nodes.}
%	\label{fig:KLd_vs_i_random_sparse_cov_cholesky_n250_linear}
%\end{figure}
%\begin{figure}[ht]
%	\centering
%	\includegraphics[width=.8\linewidth]{KLd_vs_i_random_sparse_cov_different_sqrts_n100.eps}
%	\caption{KL divergence between the distribution of the random vector $\uX$ and the model distribution after the $i$-th step of the cascade approximation v.s. the index of the cascade trees, $i$ for a graph with 100 nodes using different decompositions.}
%	\label{fig:KLd_vs_i_random_sparse_cov_different_sqrts_n100}
%\end{figure}

\subsection{The Oahu solar measurement grid dataset}
In this Example, the covariance matrix is calculated based on datasets presented in \cite{APSIPA2014}.
The Oahu solar measurement grid dataset is obtained from the National Renewable Energy Laboratory (NREL) website \cite{NREL}.
This dataset consists of $19$ sensors ($17$ horizontal sensors and two tilted sensors). 
For this dataset we normalized using standard normalization method and the zenith angle normalization method \cite{APSIPA2014}\footnote{See \cite{APSIPA2014} for more detailed description of dataset and other details about the normalization methods for the solar irradiation covariance matrix.}. From the data obtained from these $19$ solar sensors at the island of Oahu, we computed the spatial covariance matrix during the summer season at 12:00 PM averaged over a window of 5 minutes.

\begin{figure*}[t]
%	\centering
	\hspace{-1.6cm}
	\includegraphics[width=1.2\linewidth]{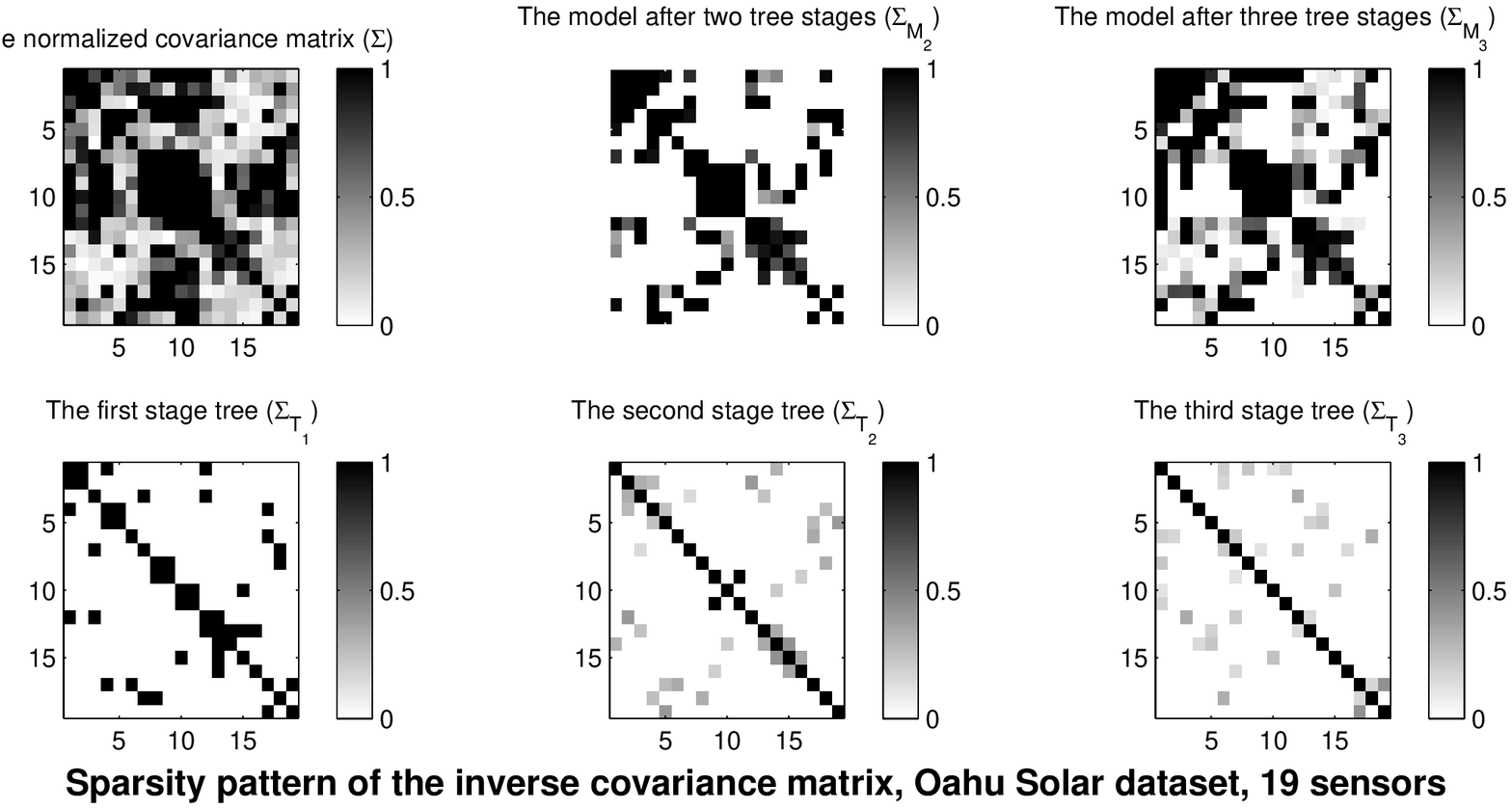}
	\caption{Gray scaled, sparsity pattern for the inverse of the covariance matrix generated using the Oahu solar measurement grid dataset. {\bf Top left:} Original normalized covariance matrix, {\bf Bottom left:} first stage tree approximation and first model. {\bf Top middle:} second approximated model, {\bf Bottom middle:} second stage tree approximation. {\bf Top right:} third approximated model, {\bf Bottom right:} third stage tree approximation.}
	\label{fig:sparsity_pattern_random_sparse_cov_n19_Oahu}
\end{figure*}

\begin{figure}[ht]
	\centering
	\includegraphics[width=1\linewidth]{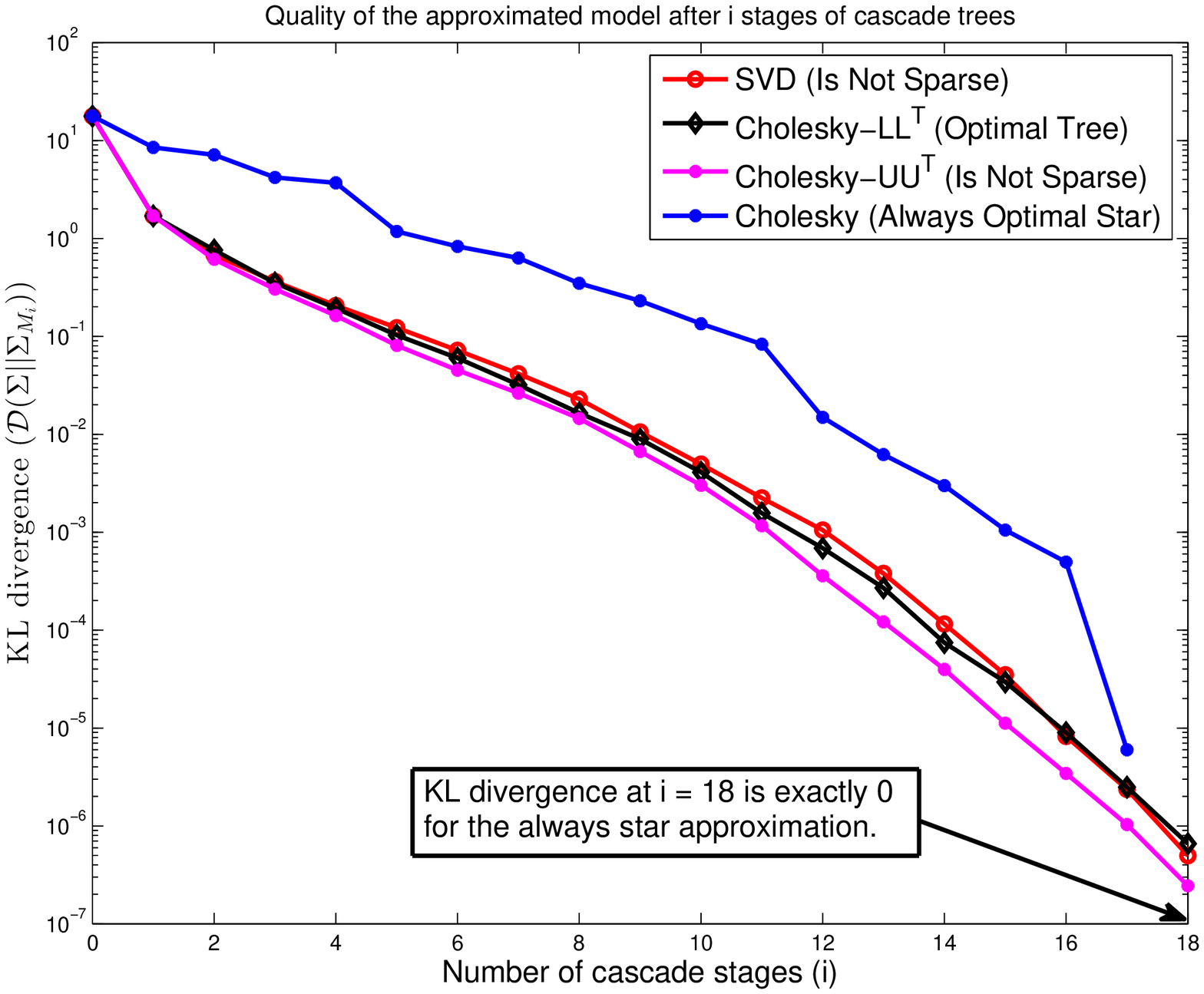}
	\caption{KL divergence between the distribution of the random vector $\uX$ and the model distribution after the $i$-th step of the cascade approximation v.s. the index of the cascade trees, $i$, for the island of Oahu solar data using different decompositions.}
	\label{fig:KLd_vs_i_random_sparse_cov_different_sqrts_n19_Oahu}
\end{figure}

Figure \ref{fig:sparsity_pattern_random_sparse_cov_n19_Oahu}  plots the gray scaled, sparsity pattern for the inverse of the Oahu solar measurement grid covariance matrix and various approximations of it. 
The top left plot shows the inverse of the original normalized covariance matrix while the bottom left plot indicates the first stage of the cascade tree approximation or the optimal Chow-Liu approximated model. 
The top middle plot shows the inverse of the second approximated model while the plot on the bottom middle indicates the inverse of the second stage tree approximation. The plot on the top right indicates the sparsity pattern of the inverse of third approximated model, while the ottom right plot shows the third stage tree approximation.

Figure \ref{fig:KLd_vs_i_random_sparse_cov_different_sqrts_n19_Oahu} plots the log-scaled KL divergence between the distribution of the random vector $\uX$ and the distribution of the model distribution after the $i$-th step of the cascade trees approximation with respect to the number of cascade trees transformation that are used in the approximation, $i$. This figure compares the performance of the proposed cascade trees approximation with different decomposition choices with the optimal star tree approximation. 
Looking only at the KL divergence we can easily see that using the greedy algorithm presented in section \ref{sec:alg} clearly has a better performance when we only have small number of cascade stages comparing to the star tree structure. On the other hand, running the cascade tree framework using the star tree approximation at each stage for $18$ stages, the KL divergence goes the zero.
This figure also plots the result of the cascade trees framework with different decompositions such as the Cholesky $\bL \bL^T$ (keep the sparsity), the Cholesky $\bU \bU^T$ (does not keep the sparsity) and the SVD. From figure \ref{fig:KLd_vs_i_random_sparse_cov_different_sqrts_n19_Oahu} we see that three of the decomposition transformations perform similarly with the star decomposition transformation performing the worse. If we compare the Chow-Liu tree to a cascade of two trees/ three trees the KL divergence decreases by respectively  more than 50\%/ 80\%. By using the Chow-Liu algorithm to produce trees and then using the Cholesky factorization in general, this algorithm performs well as the KL divergence decreases relatively quickly. However, by using the star network systematically on all nodes except one we can guarantee that the cascade algorithm converges to the model after $n-1$ steps.

\noindent {\bf Remark:} This greedy algorithm is a new way to decompose covariance matrices. We want to use the Chow-Liu algorithm since it results in the KL divergence initially decaying faster. However if we use star network, the KL divergence goes to zero after at most $n-1$ steps.

%figure \ref{fig:fig_diagonal_second_stage}
%
%\begin{figure}[ht]
%	\centering
%	\includegraphics[width=.7\linewidth]{fig.eps}
%	\caption{KL divergence v.s. the number of cascade trees for a randomly generated covariance matrix. This figure also plots the KL divergence for the second step independent (i.i.d.) approximation.}
%	\label{fig:fig_diagonal_second_stage}
%\end{figure}

\section{conclusion}
\label{sec:con}

In this paper, we look at the graphical model as a transformation and introduce a general framework to do model approximation for graphical models. This new framework, which we call the cascade trees framework, approximates a complex, hard to compute model with a cascade of simpler, more efficient tree models, that can be easily computed. To compute the optimal tree approximation at each stage of the cascade trees framework, we used the Chow-Liu algorithm.
In the computation of cascade tree framework we look at the best possible decomposition methods and we defined an important quantity, the symmetric CAM. The symmetric CAM allows us to use this cascade tree framework by creating a residual correlation matrix at each step.  The residual correlation matrix can be viewed as the remaining part of the original correlation matrix not approximated by previous iterations.  Here we used a backward construction method.
For the proposed cascade trees algorithm, the algorithm we picked uses the Cholesky lower diagonal decomposition of the covariance matrix. This choice of decomposition is favorable since it preserves the sparsity pattern of the inverse covariance matrix.
We present results that guarantees the convergence of the proposed model approximation using the cascade of tree decompositions. We confirm those results using the examples provided in the simulation section where we look at synthetic and real data and compare the performance of the propose framework by comparing KL divergences.

In future research, a more generalized case than cascade of tree models can be considered where instead of tree approximation at each iteration we look at simple, easy to compute non-tree models, such as ring models.
Here we focused on a backwards cascade model. We are currently also considering forward cascade models. We are also looking more deeply at the convergence of these cascade tree algorithms.

\section*{Acknowledgment}
This work was supported in part by NSF grant ECCS-1310634, the Center for Science of Information (CSoI), an NSF Science and Technology Center, under grant agreement CCF-0939370, and the University of Hawaii REIS project.

\bibliographystyle{IEEEbib}
\bibliography{refs}

\appendices

\section{Proof of Theorem \ref{thm:thm5}}
\label{apx:apx1}

The proof is by construction.  At each step we use a star approximation graph (a tree with one node connected to all other nodes). The Cholesky factorization is used to ensure that the sparsity pattern occurs in the inverse Cholesky matrix.
%	We proof by construction. We use the star approximation (graph with the star node having n-1 edges and all other nodes connected just to the star node), at each iteration of the cascade tree approximation algorithm. Moreover, we use the Cholesky decomposition to keep the sparsity pattern in the inverse of the Cholesky decomposition.

The first star approximation structure is constructed such that all the other nodes are connected to the first node. Then, the Cholesky decomposition is computed where the inverse Cholesky decomposition preserves the star structure. 
This particular construction causes node $1$ to become disconnected from the rest of the graph (construction of $\bDelta_1$). 

A formal, algebraic proof for this claim is given as follow. 
We generally partition the covariance matrix $\bSigma$ as
\begin{equation*}
	\bSigma =
	\begin{bmatrix}
		\bSigma_{1} & \bSigma_{12}\\
		\bSigma_{21}& \bSigma_{2}
	\end{bmatrix}
	.	
\end{equation*}
According to the covariance selection rules from Dempster theorem \cite{dempster}, the model covariance matrix has the same coefficients as the covariance matrix, $\bSigma$, at non-zero places in the inverse. For the sake of notation simplicity, let $\bSigmaT$ denotes the first stage model covariance matrix. For the proof of this part, we want the model inverse covariance matrix to have zeros at the block position of $\bSigma_2$. Thus, the model covariance matrix, $\bSigmaT$, has the following partitioned covariance matrix
\begin{equation*}
\bSigmaT =
\begin{bmatrix}
\bSigma_{1} & \bSigma_{12}\\
\bSigma_{21}& \bSigmaT_{2}
\end{bmatrix}
.	
\end{equation*}

Using matrix inversing lemma \cite{matbook} we have
\begin{equation*}
\bSigmaT^{-1} =
\begin{bmatrix}
\bSigma_{1}^{-1} + \bSigma_{1}^{-1} \bSigma_{12} \bSigmaT_{2|1}^{-1} \bSigma_{21} \bSigma_{1}^{-1} & -\bSigma_{1}^{-1} \bSigma_{12} \bSigmaT_{2|1}^{-1}\\
- \bSigmaT_{2|1}^{-1} \bSigma_{21} \bSigma_{1}^{-1}& \bSigmaT_{2|1}^{-1}
\end{bmatrix}
,
\end{equation*}
Where Schur compliment,
$$\bSigmaT_{2|1} = \bSigmaT_{2} - \bSigma_{21} \bSigma_{1}^{-1} \bSigma_{12}$$
is the conditional covariance and is diagonal.

The inverse of model covariance matrix can be factor as 
$$\bSigmaT^{-1} = \bQ^T \bQ$$
where $\bQ$ is the inverse of the lower tridiagonal Cholesky decomposition of $\bSigmaT$ and can be partitioned as
\begin{equation*}
\bQ =
\begin{bmatrix}
\bQ_1   & {\bf 0}\\
\bQ_{21}& \bQ_{2}
\end{bmatrix}
,	
\end{equation*}
where $\bQ_1$ and $\bQ_2$ can be computed from the following
$$\bSigma_{1}^{-1} = \bQ_1^T \bQ_1,$$
$$ \bSigmaT_{2|1}^{-1} = \bQ_2^T \bQ_2,$$
and $\bQ_{21}$ as follow
$$\bQ_{21} = - \bQ_2 \bSigma_{21} \bSigma_{1}^{-1}.$$

Computing $\bQ \bSigma \bQ^T$, we have
$$(\bQ \bSigma \bQ^T)_{1} = \bQ_1 \bSigma_1 \bQ_1^T = \bI$$
and
\begin{align*}
	(\bQ \bSigma \bQ^T)_{12} & = \bQ_1 \bSigma_{12} \bQ_2^T + \bQ_1 \bSigma_1 \bQ_{12} \\
	& = \bQ_1 ( \bSigma_{12} - \bSigma_1 \bSigma_1^{-1} \bSigma_{12}) \bQ_2^T  \\
	& = \bQ_1 ( {\bf 0}) \bQ_2^T = {\bf 0},
\end{align*}
and
\begin{align*}
(\bQ \bSigma \bQ^T)_{2} & = \bQ_{21} \bSigma_1 \bQ_{12} + \bQ_2 \bSigma_{21} \bQ_{12} \\
& + \bQ_{21} \bSigma_{12} \bQ_2 + \bQ_2 \bSigma_2 \bQ_2^T \\
& = \bQ_2 \bSigma_{21} \bSigma_{1}^{-1} \bSigma_1 \bSigma_{1}^{-1} \bSigma_{12} \bQ_2^T  \\
& - \bQ_2 \bSigma_{21} \bSigma_{1}^{-1} \bSigma_{12} \bQ_2^T \\
& - \bQ_2 \bSigma_{21} \bSigma_{1}^{-1} \bSigma_{12} \bQ_2 + \bQ_2 \bSigma_2 \bQ_2^T \\
& = - \bQ_2 \bSigma_{21} \bSigma_{1}^{-1} \bSigma_{12} \bQ_2 + \bQ_2 \bSigma_2 \bQ_2^T \\
& \stackrel{(a)}{=} \bQ_2 ( \bSigma_2 - \bSigmaT_2 + \bSigmaT_2 - \bSigma_{21} \bSigma_{1}^{-1} \bSigma_{12}) \bQ_2^T  \\
& = \bI + \bQ_2 ( \bSigma_2 - \bSigmaT_2 ) \bQ_2^T  ,
\end{align*}
where (a) is true since we add and subtract $\bSigmaT_2$. 

All diagonal coefficients of $\bQ_2 ( \bSigma_2 - \bSigmaT_2 ) \bQ_2^T$ are zeros, since according to the covariance selection theorem \cite{dempster}, diagonal coefficients of $\bSigma_2$ and $\bSigmaT_2$ are equal, and $\bQ_2$ is diagonal.

\noindent {\bf Remark:} Note that all diagonal coefficients of $(\bQ \bSigma \bQ^T)_{2}$ are ones since all diagonal coefficients of $\bQ_2 ( \bSigma_2 - \bSigmaT_2 ) \bQ_2^T$ are zeros.

Overall, the $\bQ \bSigma \bQ^T$ (symmetric CAM) has the following structure
\begin{equation*}
\bQ \bSigma \bQ^T =
\left[\begin{array}{c|cccc}
\bI     &        & {\bf 0} &        \\ \hline
& 1      & \boxdot & \cdots &\boxdot \\ 
{\bf 0} & \boxdot & 1      & \ddots & \vdots \\  
 & \vdots & \ddots &   \ddots   &\boxdot \\
&\boxdot & \ldots & \boxdot &    1   \\ 
\end{array}\right]
.	
\end{equation*}
For the star tree approximation we have
\begin{equation*}
\bDelta_1 = \bQ \bSigma \bQ^T =
\left[\begin{array}{c|ccc}
1      & 0       & \cdots & 0       \\ \hline
0      & 1       & \boxdot& \boxdot \\ 
\vdots & \boxdot & \ddots & \boxdot \\  
0      & \boxdot &\boxdot & 1\\
\end{array}\right]
.	
\end{equation*}

Repeating this construction process recursively, at the $i$-th iteration, node $i+1$ to node $n$ are all connected to the $i$-th node in the $i$-th star approximation structure and thus, the $i$-th node become disconnected from the rest of the graph. Repeating this procedure for $n-1$ iterations, we get $\bDelta_{n-1} = \bI$ which translates to zero model approximation error. 

Note that, we can also optimize the choice of the star tree by minimizing the KL divergence by exhaustively searching over all the $n$ possible star structures at each step of the cascade tree approximation algorithm.

\section{Proof of Theorem \ref{lem:diag1}}
\label{apx:apx2}

Our goal is to show that all diagonal coefficients of the symmetric CAM are equal to one.
For simplicity and without losing generality of the proof we can assume that the permutation matrix is identity, i.e. we start with an appropriate ordering that satisfies theorem \ref{lem:permu} conditions. Let $\bQ$ be the inverse Cholesky factorization of the tree model covariance matrix. Let us also assume that the tree model is connected. With these assumptions, matrix $\bQ$ has one non-zero coefficients in the first row (on the diagonal) and exactly two non-zero coefficients in each other row (one on the diagonal and one on its left), i.e. $\forall \; j<i\leq n, \;\; q_{ji} \neq 0 \quad\textnormal{and}\quad q_{ii} \neq 0$. This is true since $\bQ$ is a lower triangular matrix that preserves tree structure.

We can right the symmetric CAM as follow
\begin{align*}
\bQ \bSigma \bQ^T & = \bQ (\bSigma - \bSigmaT + \bSigmaT) \bQ^T \\
& = \bI + \bQ (\bSigma - \bSigmaT) \bQ^T.
\end{align*}
Note that, the difference $\bSigma - \bSigmaT$ has zeros at positions $ji$ and $ij$ where $q_{ji} \neq 0$.
To prove this lemma, we need to show that $(\bQ (\bSigma - \bSigmaT) \bQ^T)_{ii} = 0$ or equivalently we need to show that $i$th row of $\bQ (\bSigma - \bSigmaT)$ times the transpose of the $i$th row ($i>1$) of $\bQ$ is 0.
There are only two non-zero coefficients in  the $i$th row of $\bQ$, at positions $ii$ and $ji$. Thus we only need to compute $(\bQ (\bSigma - \bSigmaT))_{ii}$ and $(\bQ (\bSigma - \bSigmaT))_{ji}$.
It is easy to see that $(\bQ (\bSigma - \bSigmaT))_{ii} = 0$. For $(\bQ (\bSigma - \bSigmaT))_{ji}$, we have
$$[0 \ldots 0 \; q_{ji} \; 0 \ldots 0 \; q_{ii} \; 0 \ldots 0 ] [\boxdot \ldots \boxdot \; 0 \; \boxdot \ldots \boxdot \; 0 \; \boxdot \ldots \boxdot]^T= 0.$$
This equality holds since $(\bSigma - \bSigmaT)_{ji} = 0$. Thus, $\forall i \leq n, \;\; (\bQ (\bSigma - \bSigmaT) \bQ^T)_{ii} = 0$ which results in this lemma.

\end{document}